\documentclass[aps,prl,twocolumn,superscriptaddress]{revtex4-1}
\usepackage{amsmath}
\usepackage{graphicx}
\usepackage{braket}
\usepackage{lipsum}
\usepackage{setspace}
\usepackage[usenames, dvipsnames]{color}
\graphicspath{{figures/}{}}
\usepackage[dvipsnames]{xcolor}
\newcommand{\red}[1]{\textcolor{black}{#1}}
\newcommand{\redT}[1]{\textcolor{black}{#1}}
\newcommand{\blue}[1]{\textcolor{black}{#1}}
\newcommand{\gray}[1]{{}}
\newcommand{\purple}[1]{\textcolor{black}{#1}}
\newcommand{\grey}[1]{{}}
\begin{document}
\title{Dual-Resonator Kinetic-Inductance Detector for Distinction between\\ 
Signal and 1/f Frequency Noise }
\author{N. Foroozani}
\thanks{Requests for materials should be addressed to the corresponding author, KDO (email: kosborn@umd.edu), or NF (email: nforouzani@lps.umd.edu).}
\affiliation{Laboratory for Physical Sciences, University of Maryland, College Park, MD 20740, USA}
\affiliation{Department of Physics, University of Maryland, College Park, MD 20742, USA}
\author{B. Sarabi}
\affiliation{Laboratory for Physical Sciences, University of Maryland, College Park, MD 20740, USA}
\author{S. H. Moseley}
\thanks{Current Institution: Quantum Circuits, Inc.}
\affiliation{NASA Goddard Space Flight Center, Greenbelt, MD 20771, USA}
\author{T. Stevenson}
\affiliation{NASA Goddard Space Flight Center, Greenbelt, MD 20771, USA}
\author{E. J. Wollack}
\affiliation{NASA Goddard Space Flight Center, Greenbelt, MD 20771, USA}
\author{O. Noroozian}
\affiliation{NASA Goddard Space Flight Center, Greenbelt, MD 20771, USA}
\author{K. D. Osborn}
\thanks{Requests for materials should be addressed to the corresponding author, KDO (email: kosborn@umd.edu), or NF (email: nforouzani@lps.umd.edu).}
\affiliation{Laboratory for Physical Sciences, University of Maryland, College Park, MD 20740, USA}
\affiliation{Joint Quantum Institute, University of Maryland, College Park, MD 20742, USA}
\affiliation{Quantum Materials Center, University of Maryland, College Park, MD 20742, USA}

\date{\today}

\begin{abstract}

Astronomical Kinetic Inductance Detectors (KIDs), similar to quantum information devices, experience performance-limiting noise from materials. In particular, 1/f (frequency) noise arises from Two-Level System defects (TLSs) in the circuit dielectrics and material interfaces and can be a dominant noise mechanism. Here we present a Dual-Resonator KID (DuRKID), which is designed for improved \gray{signal-to-noise (and} noise equivalent power\gray{)} relative to standard 1/f-noise limited KIDs. \red{In this study we present the DuRKID schematic, a fabricated example, first measurement results, a theoretical
model including 1/f noise, and a system noise model containing additional noise sources.}\gray{We first show the DuRKID schematic, fabricated circuit, and we follow with a description of the intended operation, first measurements, theory, and discussion.} The circuit consists of two superconducting resonators sharing an electrical capacitance bridge of 4 capacitors, each of which hosts TLSs. The device is intended to operate using hybridization of the modes, which causes TLSs to either couple to one mode or the other, depending upon which capacitor they reside in. In contrast, the signal will affect a resonator inductance, and due to mode hybridization this causes correlated frequency changes in both modes. Therefore, one can \red{better} distinguish photon signal from TLS frequency noise. To achieve hybridization, a TiN inductor is current biased to allow tuning of one bare resonator mode into degeneracy with the other. \red{Measurements show that the resonator modes hybridize as expected.} The inter-resonator coupling and unintentional coupling of the 2 resonators to transmission lines are also characterized in measurements. \red{A quantum-information-science model allows device parameter extraction from experimental data and a 1/f noise analysis with uncorrelated noise. A system noise analysis of the DuRKID, with comparisons to standard KIDs, is performed with generation-recombination noise and amplifier noise.}\gray{In one theory, based on a quantum-information-science model, we calculate the 4-port S parameters and simulate the 1/f frequency noise of the device.} The study reveals that the DuRKID can exhibit a large \gray{and fundamental} performance advantage over TLS-limited KID detectors.  

\end{abstract}
\maketitle
\section{I. Introduction}
Kinetic Inductance Detectors (KIDs) \cite{DayPK, Mazin, NoisePropGao, AIP} for astronomy contain resonators that experience noise which is similar to the noise in qubits for quantum information processing (QIP) \cite{PRL123, Oliver, NATURE, APL2008, McRae}. For KID-based millimeter and submillimeter-wave astronomical imaging systems \cite{Maloney2010, Monfardini2011, Galitzki2016} the signal band of interest contains frequency noise that appears as a strong $1/f$-like noise spectrum. To achieve the underlying device sensitivity, scan strategies and signal modulation techniques must be implemented to mitigate the influence of noise on the final observational data products  \cite{Wright1996, Maino2002, Miller2016}. More generally speaking, if unaddressed in such applications, the presence of residual low-frequency variations introduces correlated noise which can lead to increased effective noise levels and systematic artifacts that include image striping and amplitude calibration errors. While significant progress has been made over the past two decades in reducing low-frequency noise in KIDs, developing sensitive detectors with lower noise remains the primary technical challenge for many current and future astrophysical observatories \cite{omid1, omid2, omid3}.

The performance in KIDs and qubits is sometimes limited by the same defect type, that is a defect commonly named the two-level system (TLS) which resides in dielectrics and at material interfaces within the device. KIDs are perturbed in resonance frequency by the \red{illuminating} photon signal \cite{DayPK}, however, TLSs meanwhile induce $1/f$ frequency noise which causes difficulties in signal detection \cite{NoisePropGao}. Similarly, in QIP the qubit transition experiences frequency noise from TLSs \cite{PRL123, PRB015}. This noise necessitates recalibration of pulse-driven qubit logic gates \cite{PRA016, PRL0052}. Recent studies of TLS noise mechanisms for KIDs and qubits have shown the importance of a more in-depth understanding of TLSs \cite{PRL123, PRB015, APL90-2007, PRL0052, APL92-2008}. Two recent theories explain that the interaction between TLSs is responsible for $1/f$ frequency noise whereby low-frequency thermally fluctuating TLSs influence high-frequency (near-resonance) TLSs \cite {PRB91-2015, Alex1}, which in turn interact with the high-frequency superconducting mode \cite{PRL010, PRL004}. Traditionally the influence of TLSs in KIDs and qubits is minimized through material choices \cite{Steffen, Chu, Burnett, Neill, SRON} and geometry optimization \cite {AIP}. While increasing the measurement power lowers the frequency noise from TLSs, it also generates quasiparticles. This leads to the obfuscation of the KID photon signal, \red{because the signal is itself caused by quasiparticles (via illumination of the detector)}\gray{because the signal is also produced by generating quasiparticles} \cite{refforQP}.

Recognizing that practical limitations to frequency noise in KIDs remain, we propose to tackle the problem differently. Here we describe a detector circuit that can allow one to distinguish photon signal from TLS frequency noise. The circuit uses two resonators coupled with an electrical capacitance bridge, and is named a Dual-Resonator KID (DuRKID). The design allows a shift of an inductor value through a bias current. When operating as a full detector, we imagine this inductor would also shift inductance from illumination, similar to a standard KID. In the DuRKID, the inductance shift causes correlated frequency shifts of both resonator modes. However, in the same device the TLSs will cause uncorrelated frequency shifts in the resonator modes because high-frequency TLS noise will appear independently in two groups of capacitors. We use established circuit quantum electrodynamics (cQED) analysis to model the circuit, which includes TLSs as a noise source. In\gray{ the next} section \red{II} we present the DuRKID circuit schematic and fabrication details. Data from the device, including frequency tuning and hybridization, are also shown in this section. In section III, an analysis of the DuRKID is described with representative TLSs. The analysis is used to extract the dielectric loss \red{and various coupling factors} from resonator data. \red{Section IV describes a system noise analysis of a DuRKID in comparison to a standard KID. We conclude in section V.}\gray{We conclude in section IV.}

\section{II. Device design, method of noise distinction and measurement}
TLSs shift the resonator frequency, as does the photon signal, in standard KIDs. The former effectively causes a \gray{time} change in the frequency-dependent dielectric constant, and the latter causes an increase in a superconductor's kinetic inductance \cite{Mazin2010}. We propose the DuRKID, a device with two \red{specially coupled} resonator modes\red{,} in contrast to standard KIDs, to improve the noise sensitivity that is hampered by TLS-induced frequency shifts. In the DuRKID, we find that noise from TLSs should be qualitatively different from the signal such that the two are distinguishable and the effective signal sensitivity is improved.
\begin{figure}
\includegraphics[width=0.48\textwidth]{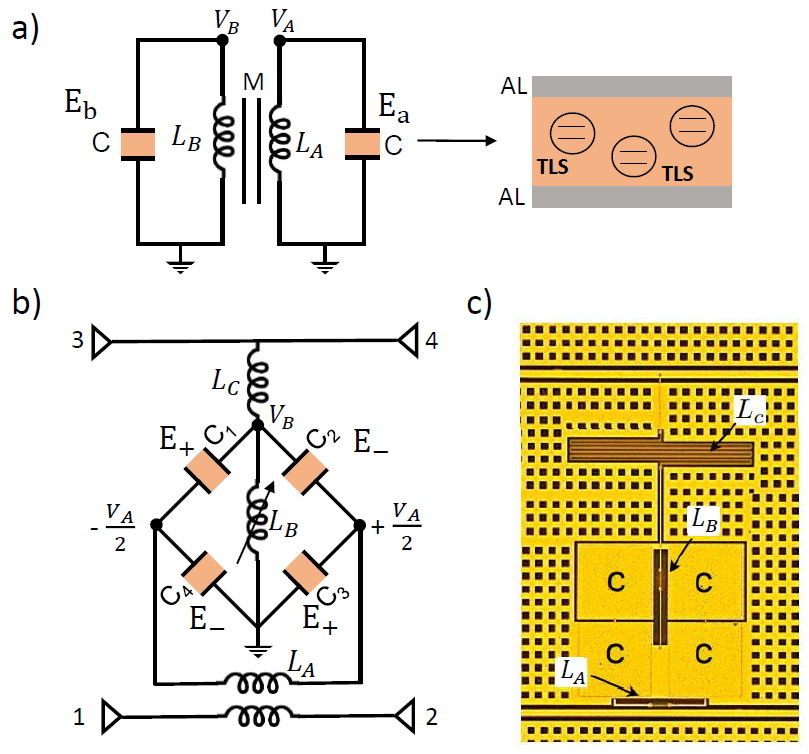}
\caption{(a, b) Resonators with two modes (an extra mode relative to a standard MKID). If the modes are far detuned (in frequency), the modes are effectively decoupled from each other and also experience different TLS noise environments. (a) Standard-coupled resonator circuit. When the two modes are hybridized, TLS noise from both capacitors will equally influence the two resonator modes: $E_a(t)=E_b(t)$ and $E_a(t)=-E_b(t)$. (b) Circuit schematic of Dual-Resonator KID (DuRKID). An electrical bridge of nominally equal capacitances ($C_1-C_4$) \red{defines the mode, as described below.} Spatial resonator modes $A$ and $B$, with voltage components $V_A$ and $V_B$, are hybridized for equal $L_A$ and $L_B$ by \gray{their} stray coupling. For the hybridized mode where $V_A(t)$ and $V_B(t)$ are in-phase, the electric field in capacitors 1 and 3 adds  $E_{+}\propto \langle a+b\rangle$, but in 2 and 4 the electric fields are out-of-phase and cancel $E_{-}\propto \langle a-b\rangle$. For the other hybridized mode, the location of constructive and destructive interference is swapped. The interference of the electric fields gives selective sensitivity of the modes to a TLS, depending on which capacitor hosts the TLS. In this schematic, a dc-bias current applied to port 3 may tune inductance $L_B$. (c) Optical image of the fabricated DuRKID. The materials include: Si substrate (black), aluminum (yellow), TiN (dark red), and silicon nitride dielectric (obscured).}
\end{figure}

We begin with a discussion of a standard-coupled resonator pair, where LC resonators are nominally the same frequency (degenerate) without coupling. This reference case is shown in Fig. 1(a), where the resonators are inductively coupled through stray mutual inductance $M$. Alternatively, they could be coupled through a capacitor for the same effect. In the schematic, each bare (uncoupled) resonance mode frequency, $f_{r, A}$ and $f_{r, B}$, is created by an individual inductor, $L_{A}$ and $L_{B}$, respectively. Without resonator coupling, a given TLS in the capacitor of each resonator will couple solely to the field of that resonator, and the resonators will act as two separate KIDs.  For each mode, one TLS causes a dispersion \red{(frequency shift): $\Delta f_{r}=(g/(2\pi))^{2}/(f_{TLS}-f_{res})$, where $f_{TLS}$ is the TLS frequency and $g$ is the coupling between the TLS and the resonator mode}. In contrast, when the resonators are set to the same bare frequency of the real device the coupling will hybridize the modes, and standard weakly coupled TLS in a capacitor will interact with both modes through the same coupling term $g$, and hence cause correlated noise in the modes. TLSs between the mode frequencies may be especially troublesome. Their frequency drift may cause a correlated drift in both modes. {\it However, \gray{a signal fed to one of the inductors} \red{illumination of the intended inductor} will also cause a correlated change in the frequencies. Thus, in a standard-coupled resonator pair hybridized modes will exhibit some correlated TLS noise that is not distinguishable from the signal.}
 
In this work, we study the DuRKID as an alternative KID. To differentiate the TLS noise from the signal, the DuRKID uses an electrical bridge for coupling between two resonator modes (see Fig. 1(b)), which adds a method to distinguish between the illumination signal and the TLS-induced frequency fluctuation noise. As in the standard-coupled resonator circuit, we have two bare (uncoupled) resonance modes $A$ and $B$, each comprised of an individual inductor, $L_{A}$ or $L_{B}$, respectively, but the capacitance bridge of the DuRKID provides four capacitors shared as an electrical bridge by the resonators. The capacitors are nominally equal such that both modes access the same nominal capacitance. The coupling according to the figure shows that mode $A$ is intentionally coupled to the transmission line with ports 1 and 2. Similarly, mode $B$ is intentionally coupled to the transmission line with ports 3 and 4. However, when hybridized, each mode is coupled to both transmission lines. To create degeneracy between these bare modes, $L_{B}$ \red{can be increased using increased dc-bias current bias}\gray{is designed to be current biased}. For example, by using \red{current injection into} port 3 we can decrease the higher mode frequency of the device shown in Fig. 1(b) because the dc-bias current enters inductance $L_B$. In this work, \gray{consider this}\red{we did not illuminate the device, but the tunable} inductor \red{has an analogous function because }\gray{analogous to the KID readout} \red{an increase in illumination could in principle increase the kinetic inductance of the same inductor. Due to this intended future functionality, we consider in theoretical analysis that this inductor is changed by illumination}.

As mentioned above, the bridge-coupled resonator pair acts differently than a standard-coupled resonator pair. A TLS in a bridge of the capacitor is equally shared by both modes before hybridization. However, once hybridized, one E-field mode amplitude is zero in 2 of the 4 capacitors. Likewise, the amplitude becomes zero in the other 2 capacitors for the other hybridized mode. {\it Thus, a given TLS will only frequency-shift (disperse) one of the hybridized modes in a bridge-coupled resonator. In contrast, a change in the frequency from \red{illuminating}\gray{incident} photons will cause a correlated shift to both modes such that \red{signal is qualitatively different than TLS noise, and thus the signal might be straightforwardly separated from this noise}.} 

\subsection{Fabrication and Measurement}
\red{We have chosen a lumped element capacitor C for our device design. Related to inductance L and capacitor C, lumped-element KIDs were proposed years ago for THz KIDs, partially because the lumped L design was found to provide a good method for collecting the created quasiparticles relative to a design that would have quasiparticles distributed throughout a cavity-length resonator\cite{LEKID}. Furthermore, a lumped-element L design has recently been used for millimeter wave astronomy \cite{DualPolLEKID}. Our fabricated design serves as a proof of concept, utilizing SiNx parallel-plate capacitors which are compact compared to capacitors that utilize the substrate or SOI-based dielectric. While lower noise capacitors could be used in the future, our fabricated design merely demonstrates one approach.}

Many DuRKIDs are fabricated together on a high-resistivity ($>$20 k$\Omega$.cm) silicon wafer. A single fabricated DuRKID is shown in Fig. 1(c). There are 3 DuRKIDs per chip, and the physical layout of a chip is shown in Fig. 2(a). Note that the center DuRKID is Device 2, and its orientation matches that of Fig. 1(b) and 1(c). Device 1 and 3 are rotated by 180 degrees relative to Device 2 and they are biased in parallel unlike Device 2. The base superconducting layer is an Al/TiN bilayer, consisting of 250 nm thick Al on top of 15 nm thick TiN. The bilayer is patterned followed by the removal of Al to leave TiN bare in certain locations to provide kinetic inductances $L_{kA}, L_{kB}$ and $L_{kC}$ as part of the total inductances $L_{A}, L_{B}$ and $L_{C}$, respectively. The dielectric layer in the capacitor bridge is a high-density silicon nitride (SiNx) film with a nominal thickness of 275 nm, deposited using PECVD \cite{pike}. Via holes in the dielectric are made using SF$_{6}$ reactive ion etching. A 250 nm sputtered Al film forms the counter electrode of the capacitor bridge. After patterning the counter electrode, excess SiN$_{x}$ is removed from most of the sample, which allows access to bonding pads on the base layer.
\begin{figure}
\includegraphics[width=0.48\textwidth]{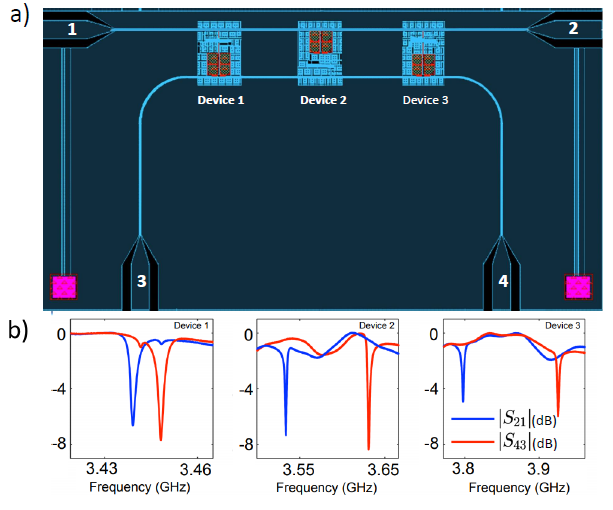}
\caption{(a) Device layout of three devices on a chip with two feedlines for each device.  One resonator of each device is tunable, using the nonlinear inductance $L_B$. The LC resonances are mainly defined by a 0.6 nH inductor and trilayer capacitors with C=3.14 pF, containing a 275 nm thick silicon nitride dielectric film. (b) Measured transmission $|S_{21}|$ and $|S_{43}|$ for devices 1, 2, and 3 at $I_b=0$.}
\end{figure}

The devices are cooled down to 20 mK in a dilution refrigerator. The transmission data $\left|S_{21}\right|$ and $\left|S_{43}\right|$ for all three devices are shown in Fig. 2(b) when $I_{b}=0$. The devices are intentionally designed with differing detunings between the two resonator
modes at zero bias current, $I_b$. The tunable resonance at zero dc bias was planned at a higher frequency than the non-tunable resonance, and this feature was realized in all three devices. The unbiased detuning between two modes for each dual resonator, Device 1-3, is measured at $\delta_{1}=9$ MHz, $\delta_{2}=97$ MHz and $\delta_{3}=128$ MHz, respectively. As intended, dc bias allows all 3 devices to reach degeneracy with bias current, and for device 3 the tuning required for degeneracy ($\sim$128 MHz) is large, but enabled by a large kinetic inductance change from the nominal inductance per square of 56 pH. The fractional frequency tuning $\delta f_{r}/f_{r}=3.5\%$ is approximately an order of magnitude higher than achieved with a magnetic-field tuned CPW resonator \cite{Vissers2015}.

\begin{figure}
\includegraphics[width=0.48\textwidth]{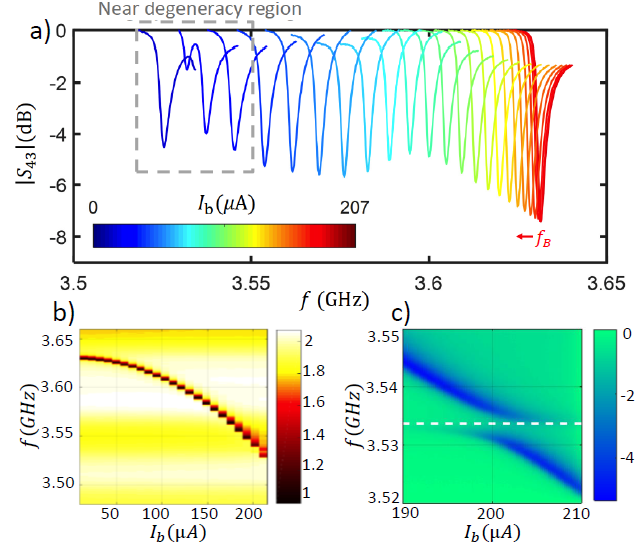}
\caption{Data on device 2. Transmission spectra, $|S_{43}|$, of tunable resonator with applied bias current up to 210 $\mu$A, measured at 20 mK. (a) Each color represents one $|S_{43}(f)|$ trace at a separate DC bias. The frequency of the resonator B, as shown by the $S_{43}$, shifts down from 3.635 GHz to 3.52 GHz, by ~120 MHz, when the minimum bias is applied for full hybridization. (b) $S_{43}(f,I_b)$ plot that shows the resonance frequency as a function of bias over the applied current range $I_b$. (c) Measured transmission $|S_{43}|$ of device B versus frequency $f_B$ and bias current $I_b$ near degeneracy shows avoided crossing due to hybridization with the fixed mode at $f_A$=3.533 GHz. From the analysis, a coupling of $\Omega_{AB}/2\pi=5.1$ MHz was obtained.}
\label{tunability}
\end{figure}

Data on device 2 transmission data $|S_{43}|$ are shown in two different plots in Fig. 3(a) and 3(b), where Fig. 3(a) shows resonance spectra in curves at different bias currents and Fig. 3(b) clearly shows resonance frequency as a function of bias current. The applied bias current shifts the higher mode toward the lower resonance and the tunable resonator disappears at a bias $I_{b,max}=210 \mu A$, indicating a change to the normal state, likely at the wiring vias. Although most of the geometric inductor length within each resonator is provided by Al, the kinetic inductance $L_{kA}$ and $L_{kB}$ comprise approximately $85\%$ of the total inductance $L_{A}$ and $L_{B}$. At temperatures much lower than the superconducting critical temperature $T_{c}$, the nonlinear response of $L_{k}$ to $I_{b}$ can be expanded as $L_{k}(I_{b})=L_{k}(0)(1+(\frac{I_{b}}{I'})^{2}+...)$, where $I'$ sets the scale of nonlinearity. By considering up to quadratic order in $L_{k}(I_{b})$, we extract $I' = 789 \mu A$ from a fit to $L_{kB}$ (See Fig. 6 in Appendix A). 

$|S_{43}|$ at different $I_{b}$ (given in Fig. 3(c)) shows an avoided crossing due to hybridization with the A mode. The operation of the device is intended for full hybridization, which occurs at $I_b \approx 202 \mu A$ in this device. From this bias current, one can simultaneously measure the two resonances using two fixed-frequency sources. These are the tones that are necessary for operating the DuRKID as intended. 

We also measured the cross-coupling from the input port of one transmission line to the output port of the other, $S_{41}$ and $S_{23}$. The transmission $|S_{41}|$ for degenerate hybridized modes is on the order of -10dB (not shown), indicating some internal loss in the resonators but also good coupling of the resonators to their intended transmission lines.

The capacitors of the bridge are carefully designed in layout to be equal. As a result, the mutual inductance is likely the dominant \red{stray} coupling mechanism. We note that the splitting from mode coupling must be larger than other line widths for the device to operate as intended. \red{To further analyze the data quantitatively, we fit the transmission data, $S_{21}$ and $S_{43}$, to the two resonator model, where the fitting procedure is explained in Appendix A.} From Fig. 3(c) we obtained the inter-resonator coupling of $\Omega_{AB}/2\pi=5.1$ MHz. This coupling agreed, within a factor of 2, with the simulated value. We then use this as a fixed parameter in the device model for a later part of the fitting procedure. We fit both standard transmission data, $S_{21}$ and $S_{43}$, at the same time for the start of the fit procedure. From this, we extract the internal Q-factor of 2750 and 3100 for resonators $A$ and $B$, respectively. When viewed as a material loss tangent, tan$\delta=1/Q_{i}=(3.4\pm0.2)\times 10^{-4}$ and the value of this loss matches our expectation for the recipe of PECVD SiN$_{x}$ that we used. This loss tangent is lower than AlO$_{x}$ barriers in large-area Josephson junctions by approximately an order of magnitude \cite{Martinis2005PRL, pike}. In other work, silicon-on-insulator (SOI) is used for higher Q-factor resonators \cite{crystal, Ed}. \red{This dielectric type is advantageous for KIDs due to its low loss characteristics, resulting in low noise from TLSs.}

\blue{We observed unintended leakage (cross talk) between the two different feedlines at $|S_{14}|\sim 10^{-4}$. This small value was measured with resonators detuned and it corresponds to the stray coupling at the frequency of either the A or B resonator mode. From this, we extracted the couplings of the resonators to the unintended feedlines as  $\kappa_{ax} = \kappa_{bx} \sim 20$ kHz, where we define $\kappa_{ix}$ in the next section (also see Appendix B for details).} The leakage will generally produce a negligible change to measurements using two feedlines. However, the two hybridized modes can also be read out with one feedline for the sake of simplicity.

\red{It is worth noting that the DuRKID can be multiplexed. One advantage of standard KIDs is that they can be read out at different frequencies using frequency-division multiplexing using one feedline. This multiplexing has enabled $\approx$1000-pixel KID arrays \cite{ThousandPixel}. However, in the present DuRKID design, dc-bias current must also be applied and it is presently applied through an RF feedline. To achieve the intended biasing for a $\approx$1000-pixel DuRKID array, one can use 32 bias lines that are separate from the feedlines, to bias a column of DuRKIDs using a parallel circuit. The microwave measurement of the full array can then occur in 32 measurements in time from one feedline, where each uses biases set for one row of DuRKIDs. 
}

\section{III. C-QED model of two resonators coupled to the same TLS}
\subsection{A. Theoretical Model}
For a theoretical model of the DuRKID, we use a standard quantum information science method (c.f. Ref. \cite{Gardiner}). Fig. 4(a) shows the system diagram for the model, which has two resonator modes. Per the layout, we describe each resonator mode with its own two-port transmission line. Both modes are coupled to a TLS noise source. Only one TLS and \red{its} coupling is shown, but the model generally has many TLSs \red{with} different couplings to one hybridized resonator mode, \red{and the hybridized mode is generally created from}\gray{due to their symmetry created by} inter-resonator mode coupling $\Omega_{AB}$.  $\kappa_{A}$ and $\kappa_{B}$ are the coupling rates of resonators $A$ and $B$, respectively, to their intended 2-port transmission line. The coupling rates of the resonators to the other (unintended) transmission lines are represented by $\kappa_{Ax}$ and $\kappa_{Bx}$, respectively. In this analysis, we focus on TLSs that are far enough from resonance to not be saturated by the drive field. A coupled TLS pair could cause different qualitative noise spectra (c.f. Ref. \cite{PRB015}), but the method of protection is qualitative \red{such that}\gray{and thus} the interaction of the pair within a capacitor should not matter.
Generally, we consider that the resonators are coupled to the \textit{i}-th TLS with a resonance coupling constant of $g_{i}$. We represent TLSs with a spin operator $\sigma_{i}^{z}$, using the analogy between TLS theory and a two-state system (with pseudo spin-$\frac{1}{2}$). The Hamiltonian for the system interacting with two transmission lines then has the form
\begin{equation}
\begin{aligned}
H_{sys} &= \hbar\omega_{A}a^{\dagger}a+\hbar\omega_{B}b^{\dagger}b+\sum_{i=1}^{N}\varepsilon_{i}\sigma_{i}^{z}+\sum_{j=1}^{M} \varepsilon_{j}\sigma_{j}^{z}\\
& +\hbar\varOmega_{AB}(b^{\dagger}a+a^{\dagger}b)\\
& -i\hbar\sum_{i=1}^{N}g_{i}^{(n)}(\sigma_{i}^{+}(a+b)+\sigma_{i}^{-}(a^{\dagger}+b^{\dagger}))\\
 & {-i\hbar\sum_{j=1}^{M}g_{j}^{(m)}(\sigma_{j}^{+}(a-b)+\sigma_{j}^{-}(a^{\dagger}-b^{\dagger})).}
\end{aligned}
\end{equation}
\begin{figure*}[htp]
\includegraphics[width=0.8\textwidth]{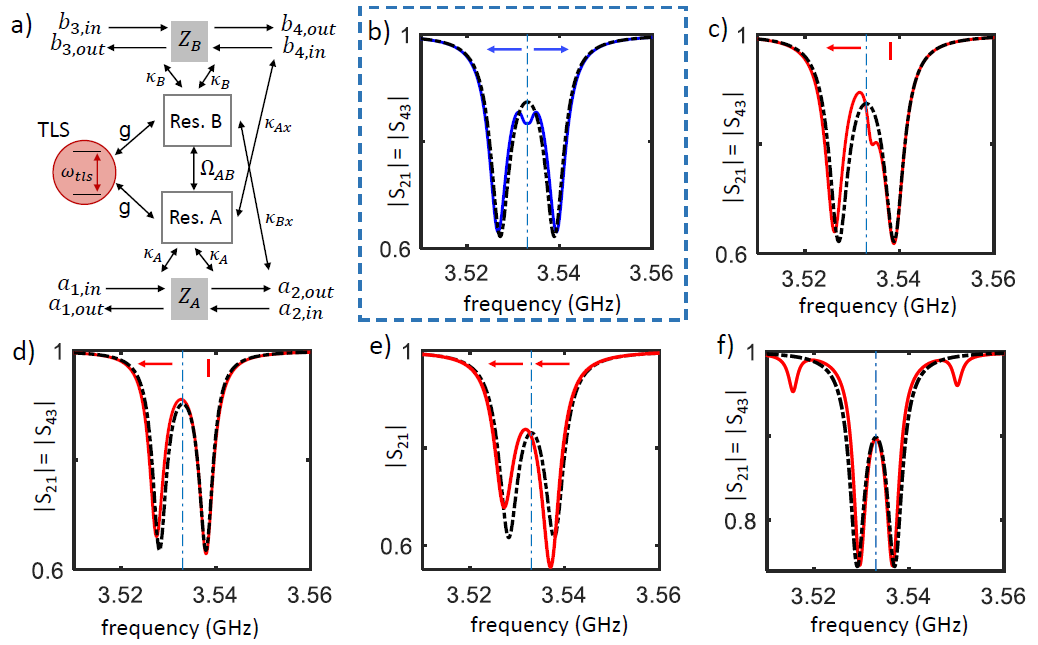}
\caption{(a) Theoretical model for the DuRKID, including 2 resonators, 4 ports (fully shown), and TLSs (only 1 TLS shown). In this simulation, the coupling parameters are $\kappa_{A}/2\pi=\kappa_{B}/2\pi=2$ MHz, $\gamma_{TLS}/2\pi=2$ MHz, $\varOmega_{AB}/2\pi=5$ MHz and $g/2\pi=1$ MHz. Resonator A is intentionally coupled to the transmission line of port (1) and (2), and resonator B to port (3) and (4). (b-f) Transmission analysis for $(1\rightarrow2$ or $3\rightarrow4)$ with $Z_{A}, Z_{B}\rightarrow0$ (zero impedance between transmission line halves). Arrows (red or blue) show resonator mode frequency (solid curve transmission minima) change due to the TLS, relative to the mode frequency without the TLS (dashed curve transmission minima). (b-c) Single-photon transmission spectrum for two different devices with same parameters, $(\omega_{A}/2\pi=\omega_{B}/2\pi=3.533$ GHz) with a single degenerate TLS $(\omega_{tls}=\omega_{A}=\omega_{B})$. (b) Standard-coupled resonators which are hybridized. Note that the frequency shift (dispersion) occurs equally (correlated) in both modes (blue relative to black minima, also see blue arrows). (c) DuRKID in intended (hybridized) mode with a TLS. Here the TLS leaves one of the modes undispersed. (d) DuRKID with a large drive (classical field) amplitude of $\overline{n}=50$ photons and a TLS frequency of $\omega_{TLS}/2\pi=3.540$ GHz. Here the dispersion is present, related to (c), and the TLS is symmetry-protected from saturation. Undispersed modes occur in DuRKID due to the absence of fields in certain capacitors according to the symmetry of the hybridized modes (see Figure 1(b)). (e) The spectra for the bridge-type resonator device in the presence of an added signal to one inductor (in contrast to other sub-figures that use additional TLS(s)). Here the hybridization of the modes changes as seen in different resonator transmission amplitudes. The resonator frequency modes shift downward together (as correlated modes), in contrast to the case of single-TLS perturbation (c or d). (f) Spectra with two TLSs, where there is one from each group (where each mainly influences only one of the resonator modes). This case is used in section III B, for analysis with two TLS noise sources. A TLS in each TLS group disperses each hybridized mode, but it is distinguishable from the (correlated-mode) signal. In this special case, there is no change of mode hybridization due to the way that the TLSs oppositely disperse (shift) the resonator modes.}   
 \label{optical image}
\end{figure*}This is similar to the one-cavity Jaynes-Cummings model \gray{, which unlike the standard Jaynes-Cummings Hamiltonians} \cite{Walls, Fink2008, Fink2009}, \red{except ours} includes 2 modes (with operators $a$ and $b$) with many TLSs (with raising and lowering operators $\sigma_{i}^{+}$ and $\sigma_{i}^{-}$). The \textit{i}th TLS has the energy of $\varepsilon_{i}=\frac{\hbar\omega_{TLS,i}}{2}=\sqrt{\varDelta_{i}^{2}+\Delta_{0,i}^{2}}$, where $\varDelta_{0}$ and $\varDelta$ are tunneling energy and the asymmetry energy in the double-well potential model. TLS resonance coupling to the resonator field is $g_{i}=\frac{\varDelta_{0,i}}{\varepsilon_{i}}p_{i}cos\theta_{i}\sqrt{\frac{\omega}{2\epsilon_{r}\epsilon_{0}\hbar V}}$, where $p_{i}$ is the magnitude of the dipole moment, $\theta_{i}$ is the angle between $p_{i}$ and the applied electric field $\bf E$, V is the dielectric volume, and $\epsilon_{r}$ is the relative permittivity \cite{bahman2}. Though a given TLS in a capacitor is equally shared by both modes before hybridization; after hybridization, the two-mode fields become $E_{+}{\color{black}\propto}(a+b)$ and $E_{-}{\color{black}\propto}(a-b)$, which are zero in certain capacitors as mentioned above. It becomes zero in 2 of the capacitors for one hybridized mode and similarly zero in the other 2 capacitors of the other hybridized mode. We find that this approximation is valid even for a TLS throughout the eigenmode width. For the two modes we thus effectively have only the coupling to N TLSs (group n) in capacitor pair ${C_{1}, C_{3}}$, and M TLSs (group m) in ${C_{2}, C_{4}}$. Using a standard theoretical procedure \cite {inputout} for the resonators $A$ and $B$ interacting with the input and output fields as the heat bath, the Heisenberg equations of motion can be written as:
\begin{equation}
\begin{aligned}
\frac{d}{dt}\left\langle a\right\rangle =&-i\omega\left\langle a\right\rangle =-\frac{i}{\hbar}\left\langle [a,H_{sys}]\right\rangle -(\kappa_{A}+\gamma_{A}+\kappa_{Ax})\left\langle a\right\rangle\\
	&+\sqrt{\kappa_{A}}\left\langle a_{1,in}\right\rangle +\sqrt{\kappa_{A}}\left\langle a_{2,in}\right\rangle +\sqrt{\kappa_{Ax}}\left\langle b_{3,in}\right\rangle \\
	&+\sqrt{\kappa_{Ax}}\left\langle b_{4,in}\right\rangle-\sqrt{\kappa_{A}\kappa_{Bx}}\left\langle b\right\rangle -\sqrt{\kappa_{Ax}\kappa_{B}}\left\langle b\right\rangle,
\end{aligned}
\end{equation}

\begin{equation}
\begin{aligned}
\frac{d}{dt}\left\langle b\right\rangle =&-i\omega\left\langle b\right\rangle =-\frac{i}{\hbar}\left\langle [b,H_{sys}]\right\rangle -(\kappa_{B}+\gamma_{B}+\kappa_{Bx})\left\langle b\right\rangle\\
	&+\sqrt{\kappa_{B}}\left\langle b_{3,in}\right\rangle +\sqrt{\kappa_{B}}\left\langle b_{4,in}\right\rangle +\sqrt{\kappa_{Bx}}\left\langle a_{1,in}\right\rangle\\
	&+\sqrt{\kappa_{Bx}}\left\langle a_{2,in}\right\rangle 
	-\sqrt{\kappa_{B}\kappa_{Ax}}\left\langle a\right\rangle -\sqrt{\kappa_{Bx}\kappa_{A}}\left\langle a\right\rangle,
\end{aligned}
\end{equation}	
where $a_{1,(2),in}, b_{3,(4),in}$ and $a_{1,(2),out}, b_{3,(4),out}$ are the input and output fields. Assuming a coherent-state approximation for photons in resonators $A$ and $B$ coupled to input and output fields, the transmissions $S_{21}$ and $S_{43}$ are given by $\frac{\left\langle a_{2,out}\right\rangle }{\left\langle a_{1,in}\right\rangle}$ and $\frac{\left\langle b_{4,out}\right\rangle }{\left\langle b_{3,in}\right\rangle }$, respectively. The boundary conditions which relate the input and output fields to photon annihilation in each resonator, $a_{1,(2),out}=\sqrt{\kappa_{A}}a-a_{2,(1),in}+\sqrt{\kappa_{Bx}}b$ and $b_{3,(4),out}=\sqrt{\kappa_{B}}b-b_{4,(3),in}+\sqrt{\kappa_{Ax}}a$, are used in the analysis \cite{inputout}. We assume only one input field $a_{1,in}$  (\textit {i.e.}
 $a_{2,in}=b_{4,in}=b_{3,in}=0)$ is present for $S_{21}$ and $b_{3,in}$ (\textit{i.e.} $a_{2,in}=b_{4,in}=a_{1,in}=0)$ for $S_{43}$. The details of this theoretical analysis are described in Appendix B. 

We next confine ourselves to degenerate modes $\omega_{A}=\omega_{B}=\omega_{r}$, and, for simplicity, M-group TLSs (which belong to the capacitor pair labeled with m). 
We can see from the Hamiltonian that for positive $\varOmega_{AB}$ the M TLSs are interacting with the lower frequency mode. We also reduce the calculation to equal coupling and photon decay rates in each resonator: $\kappa_{A}=\kappa_{B}=\kappa_{r}, \gamma_{A}=\gamma_{B}=\gamma_{r}$, and assume each resonator is decoupled from the other transmission line, $\kappa_{Ax}=\kappa_{Bx}=0$. In the low-temperature and low-drive power limit $(k_{B}T\ll\hbar\omega$ and $n_{ph}\ll1)$, the resonator will be asymptotically close to the ground state and we find the transmission amplitude of this four-port system as	
\begin{widetext}
$S_{43}=S_{21}=1-\kappa_{r}\frac{-i(\omega-\omega'_{r})+\kappa_{r}+\gamma_{r}}{-(\omega-\omega'_{r})^{2}+(\kappa_{r}+\gamma_{r})^{2}-(i\Omega_{AB}-\sum_{i}^{M}\frac{\tanh(\frac{\hbar\omega}{2k_{B}T})g_{i}^{2}}{-i(\omega-\omega_{TLS}^{i})+\frac{\gamma_{TLS}^{i}}{2}})^{2}-2i(\kappa_{r}+\gamma_{r})(\omega-\omega'_{r})}, \quad \quad \quad \quad \quad\quad \quad (4)$
\end{widetext}
where $\omega'_{r}=\omega_{r}-\sum_{i}^{M}\frac{\tanh(\frac{\hbar\omega}{2k_{B}T})g_{i}^{2}}{\omega-\omega_{TLS}+i\frac{\gamma_{TLS}}{2}}$. \red{For completeness, we provide in Appendix B} the transmission for TLSs interacting with the high-frequency mode (Eq. 19) and that of the standard-coupled resonator pair (Eq. 24), as well as the cross-transmission from port 3 to 2 $(S_{23})$ and from port 1 to 4 $(S_{41})$, with non-zero $\kappa_{Ax}$ and $\kappa_{Bx}$ (Eqs. 18, 19).
\begin{figure*}
\includegraphics[width=0.85\textwidth]{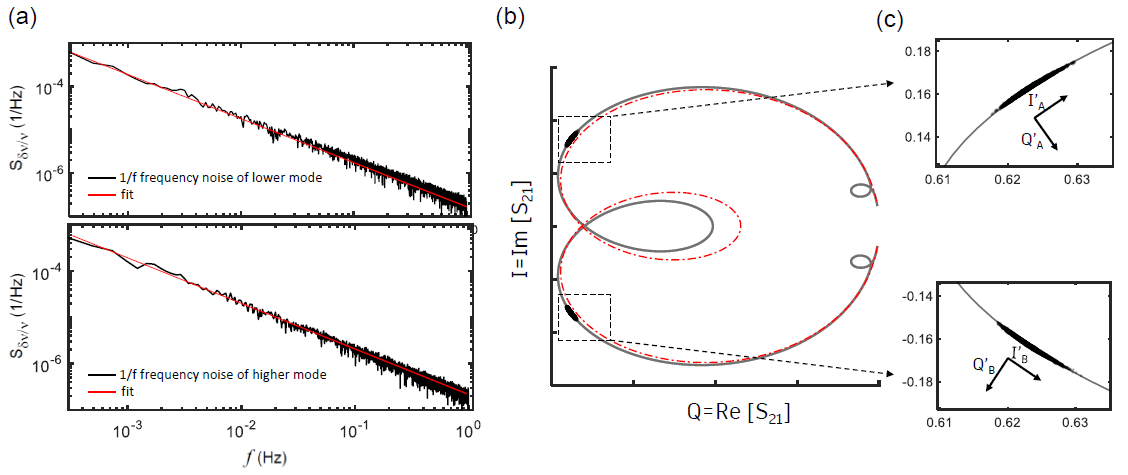}
\caption{Noise analysis for $1/f^{\alpha}\sim1/f$ TLS noise, using time-domain simulations for two variable-frequency TLSs. (a) Fractional frequency noise spectral density $S_{\delta \nu /\nu}$ for the two resonator modes. The solid line is fit to A/[f /(1Hz)] + B, where we define A as the $1/f$ noise density and B as the white noise density. (b) Complex transmission, Im[$S_{21}$] versus Re[$S_{21}$], for the DuRKID with two TLSs (black line), where there is one TLS from the m and n groups, which according to the design implies that each TLS couples to only one mode. The red dashed line is shown for the transmission without TLSs (this was numerically obtained with zero TLS coupling, which is equivalent). The solid and dashed lines can be compared to transmission magnitude versus frequency, Fig. 4(f), where there are 4 minima in the case of coupled TLSs. Transmission time-domain simulation at 2 fixed input frequencies (black points in 2 dashed boxes). The 2 sets of accumulated points are further apart in I=Im$[S_{21}]$ than the Q=Re$[S_{21}]$ minima (solid line) because the input frequencies for TLS noise analysis are chosen to be equal to the case without coupled TLS, and as Fig. 4(f) shows: the resonance minima for the case with TLS (solid line of Fig. 4(f)) are closer in frequency than the case without TLS coupling (dashed line Fig. 4(f)). (c) Zoomed view of transmission points accumulated in time for fixed input frequencies (black points). The transmission for various frequencies is shown as a grey line. (upper panel) $I'_A$ quadrature is defined as the tangent of the transmission versus frequency on the 1st mode. (lower panel) $I'_B$ quadrature is defined as the tangent of the transmission versus frequency on the 2nd mode. TLS noise in the resonator modes is uncorrelated due to separated noise sources, unlike the single-mode frequency-quadrature signal from a standard KID.}   
 \label{optical image}
\end{figure*}
\subsection{B. Theoretical Results}
In order to investigate the effect of TLS noise on the two resonator modes, we use the analytical results derived from Eq. (4) for both transmission lines in the limit of a weakly coupled TLS where $g	\ll \varOmega_{AB}$. The system we consider is shown in Fig. 4(a). Figs. 4(b) and (c) show the spectra of transmission $|S_{21}|$ for a TLS with the uncoupled modes in the standard-coupled design and the DuRKID, respectively. As shown, a given TLS in the DuRKID generally only shifts one of the two hybridized modes (Fig. 4(c)), as expected. This is true even for a TLS that is detuned from the coupled modes (related to hybridized mode symmetries). In contrast, a TLS in either capacitor in the standard degenerate design will couple with equal strength to the two modes and cause a frequency shift to both (see Fig. 4(b)). The difference is evident from the TLS-resonator interaction term in Eq. (1) which differs from the standard-coupled resonators (see Eq. 24 in the Appendix). Thus the DuRKID has single-mode noise for a single TLS, whereas a standard-coupled TLS has correlated-mode noise (see Appendix B for more data on the off-resonance TLS case). We also consider the case of a strong readout tone from port 1 or 3 (see also Appendix A). In Fig. 4(d) the transmission spectra are shown in the presence of strong pump drive with $H_{d}=i\hbar\sqrt{\kappa_{A}}A(t)a^{\dagger}$, where $\left\langle a_{1,in}(t)\right\rangle =A(t)$ and $\left\langle a_{2,in}(t)\right\rangle =\left\langle b_{3,in}(t)\right\rangle =\left\langle b_{4,in}(t)\right\rangle =0$. As shown, a given TLS in the bridge capacitor design, even in the presence of the strong pump drive, only shifts one of the two hybridized modes. Figure 4(e) shows the simulated signal on both resonance modes in the bridge resonator design viewed from transmission spectra $1\rightarrow2$ (with $Z_{A}, Z_{B}\rightarrow0$ ). Specifically, the solid line shows the transmission with a signal that increases $L_{B}$, as expected for a KID. We see relative to the unperturbed transmission (the dashed line) that the transmission spectra have a shift in both resonance modes to a lower frequency, as expected from earlier arguments. In addition, the relative sizes of the observed transmission notches are changed due to the way the modes are changed qualitatively -- one mode is more A-like; the other mode is more B-like.  In summary, one DuRKID mode will experience frequency shifts from one set of capacitors (m-group TLSs in equation 4) that will shift a single mode; this is distinguishable from two correlated modes caused by the signal. 

Figure 4(f) shows the transmission spectra of the DuRKID in the presence of two TLSs, one from each group (capacitor pair). Next, this arrangement with two TLSs is used to calculate the cross-spectral density with TLSs and the two noise sources. For this, we simulate the TLS-induced frequency noise of two resonator modes split by $\Omega_{AB}$, with bare resonance frequencies of $\omega_0/2\pi = 3.533$ GHz. The frequency noise spectral density $S_{\delta \nu /\nu}(f)$ is simulated with one TLS from each TLS group (m or n), by setting a $1/f$ noise spectrum on each. 

In Fig. 5(a), we show the simulated frequency-noise spectral density for resonators $A$ and $B$. TLS frequency fluctuations in turn create fluctuations \red{in} the resonator modes, according to their effective couplings. The TLS with higher energy is coupled to the higher resonator mode while the lower energy TLS is coupled to the lower mode. These conditions create uncorrelated $1/f$ frequency noise in the resonator modes (see Appendix B). Im[S$_{21}]$ versus Re[S$_{21}]$ is shown in Fig. 5(b) and (c) without any TLSs interaction, $g=0$, (red line) and the TLSs interaction (grey line).
\red{In this section (III) we analyze 2 TLSs from 2 separate capacitor groups. This theoretical analysis of the 2 TLS, one in each capacitor group, resulted in uncorrelated noise from the TLSs as expected (see Appendix B). The noise is more completely modeled in the next section, using additional noise sources which will be present, along with 
an extra expected stray capacitance.}\gray{Even though we have not analyzed more TLSs, from the principle of superposition we can expect the uncorrelated noise feature will continue for a typical TLS distribution -- TLSs noise should not be correlated in separate capacitors. This is distinct from correlated noise in the modes, which exists only where expected (see Appendix B, section 3).}
\section{IV. System Noise Analysis}
\red{Next we compare the expected noise from the DuRKID readout with that of a standard KID. In our earlier discussion, we mainly analyzed the TLS noise in terms of frequency change (and frequency quadrature in transmission).  However, a standard KID partially mitigates TLS noise by also measuring the dissipation quadrature. Thus, this analysis is important for a fair comparison.}

\begin{figure}
\includegraphics[width=0.45\textwidth]{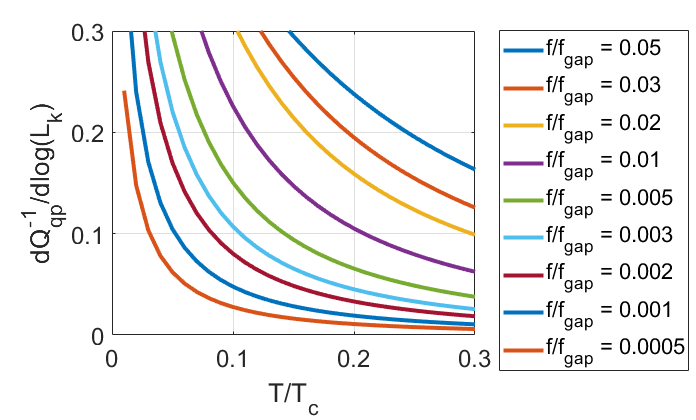}
\caption{Responsivity ratio between dissipation and frequency readout channels for a KID, as a function of reduced operating temperature and readout frequency. The noise analysis in this section is calculated using a responsivity ratio of 0.03.
}
 \label{responsivity}
\end{figure}

\red{Our model starts by specifying a dissipation to frequency responsivity (a responsivity ratio) from quasiparticles created in the illuminated inductor. Fig. \ref{responsivity} shows this responsivity ratio calculated from the Mattis-Bardeen formula as a function of the operating temperature relative to the superconducting transition temperature, and of the resonance frequency relative to the superconducting gap frequency.  For some applications, e.g. NASA's EXCLAIM mission \cite{EXCLAIM}, the reduced temperature and frequency operating point corresponds to a relatively large responsivity ratio, and a single mode KID can mitigate TLS noise fairly well. The implementation uses thin-film aluminum MKIDs operating at 168 mK and 3.5 GHz.  Furthermore, in the EXCLAIM mission $(T/T_c, f/f_{gap}) = (0.13, 0.035)$ which results in a responsivity ratio of $dQ_{qp}^{-1}/d\log(L_k) \approx 0.3$ in the KID. Here, $\log(L_k)$ is the natural log of the kinetic inductance, and $Q_{qp}$ is the internal quality factor from quasiparticles. For other circumstances, one may employ larger capacitors and also a lower resonance frequency such as a thin aluminum resonator operating at 200 mK and 0.5 GHz, which has $(T/T_c, f/f_{gap}) = (0.15, 0.005)$, giving the responsivity ratio of $\approx 0.095$.  In our noise analysis, we choose the responsivity ratio of 0.03 as a realistic experimental condition that creates $1/f$-noise limitations.}

\red{Similar to previous studies \cite{NEP1,NEP2,NEP3}, our noise model describes the Noise Equivalent Power (NEP).  Besides TLS-induced $1/f$ noise, the model includes generation-recombination (G-R) noise from quasiparticle fluctuations in the superconducting inductors \cite{deVisserThesis} and readout amplifier noise.  The signal frequency response of the detector is assumed to roll off with a time constant equal to the quasiparticle lifetime \purple{$\tau_{qp}$}. In the model, the TLS-induced $1/f$ noise is introduced through capacitance fluctuations, which captures a sufficient level of detail for the purpose at hand. For the DuRKID, we used a circuit model to evaluate the responsivity of the in-phase (\grey{imaginary}\purple{real} part) and quadrature (\grey{real}\purple{imaginary} part) of \grey{$S_{21}$}\purple{$S_{43}$} to changes in each capacitor in the circuit, as well as to changes in the inductance and dissipation (in accord with the responsivity ratio) from illumination.}

\begin{figure}
\includegraphics[width=0.47\textwidth]{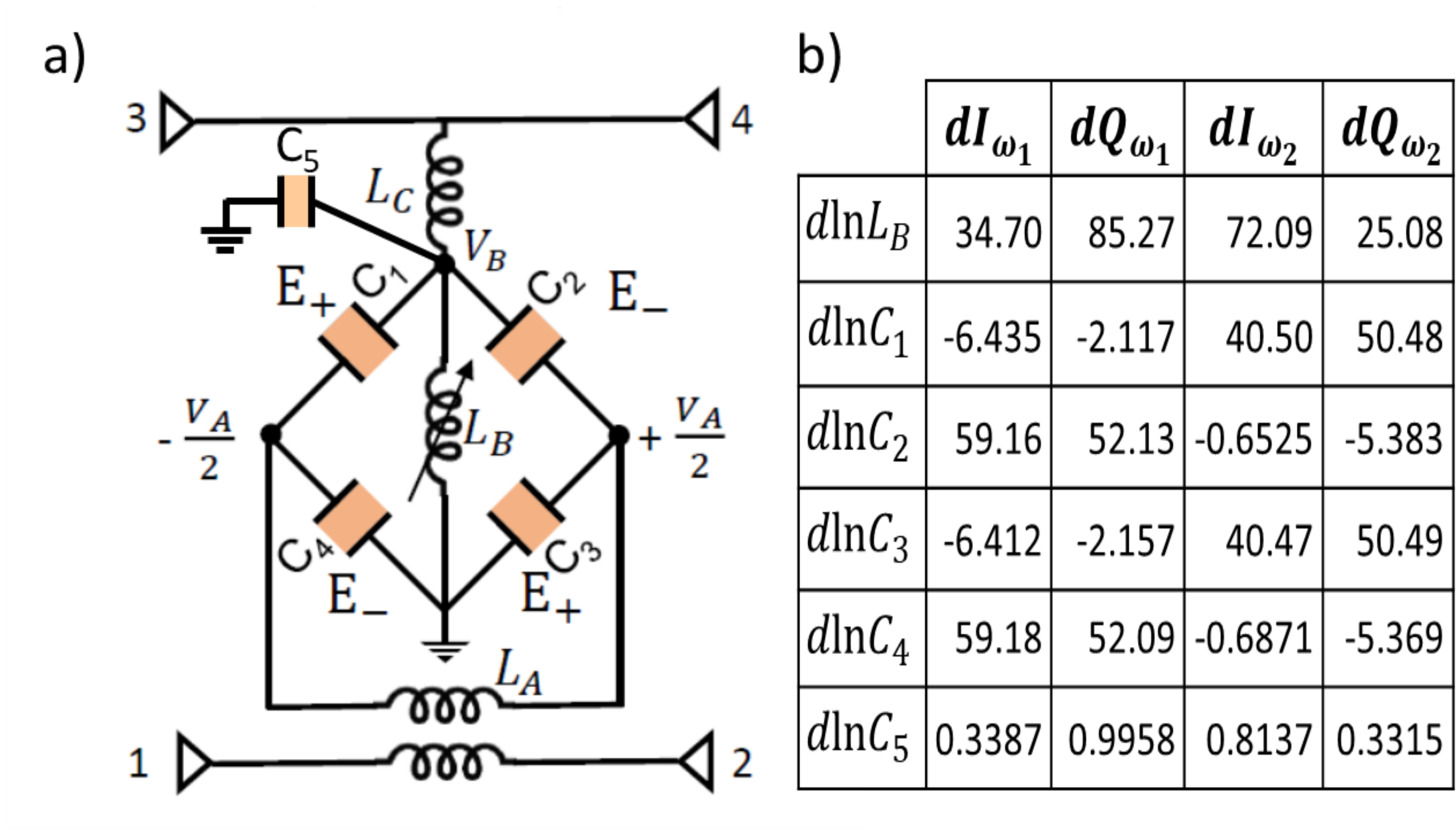}
\caption{(a) Circuit schematic for DuRKID. This schematic includes stray capacitance $C_5$, which is outside of the intended capacitive electrical bridge. (b) \grey{Cross-spectral density}\purple{Responsivity} matrix \purple{$\bf{M}$. This is the derivative of the }\grey{: $S_{1,2}$} vector \grey{response from the 4}\purple{composed of the four} DuRKID measurement channels (I-Q \purple{components of $S_{43}$} at \grey{two tone}\purple{the two resonance} frequencies) \grey{and 6 changes in the detector}\purple{with respect to the fractional change of six detector parameters}. \grey{The changes arise from}\purple{The detector parameters with fluctuations are}: the \grey{inductor signal}\purple{inductance} $L_B$ (with a dissipative response direction dictated by the responsivity ratio), and \grey{5}\purple{five} capacitances $C_1 - C_5$ (related to TLS-induced noise). \red{The size of changes is modeled using device parameters, including measured quality factors.}
}
 \label{NoiseCircuit}
\end{figure}

\redT{The model of the DuRKID for this section's noise analysis is shown in Fig. \ref{NoiseCircuit} (a). \redT{For a realistic model, we added capacitance $C_5$, which was not shown earlier in Fig. 1 (b). This represents the stray capacitance in our circuit, as it is not present in an ideal capacitance bridge.} TLSs in this capacitance will appear on the resonance mode B (before hybridization) such that $C_5$ \gray{later} introduces noise on both modes when hybridized: nominally the ($C_1$, $C_3$) mode and the ($C_2$, $C_4$) mode are affected when the detector is perturbed by $C_5$. The stray capacitance and relative capacitance changes from TLS in the analysis are estimated from the fabricated circuit geometry shown in Fig. 2 (a).}

\redT{To compute the NEP, we used the method in Sec. IIB of Ref. \cite{Stevenson}, which describes the theory of optimal linear filtering to maximize the signal-to-noise ratio for a detector with a vector output \purple{$\bf{x}$}.  In the case of a DuRKID, the detector system has a four-component vector output comprised\purple{ , for example,} of the real and imaginary parts of \grey{$S_{21}$ }\purple{$S_{43}$ }\grey{measured at two microwave probe tone frequencies}\purple{probed at the two microwave resonance frequencies} (similar to \purple{the transmission shown in }Fig. 5(c)).  The first step of the NEP calculation involves evaluating the cross-spectral density matrix \purple{$\bf{S_x }$} between the components of the signal vector \purple{$\bf{x}$} using the assumed noise sources and the responsivity of the signal to changes in circuit parameters. One such \purple{responsivity }\grey{density }matrix \purple{$\bf{M}$} is shown in Fig. \ref{NoiseCircuit} (b), where the first row represents the components from an illumination signal, which we describe for brevity as being optical. In the remaining 5 rows, we represent the components from the TLS-induced capacitance noise.} \purple{The cross-spectral density matrix is $\bf{S_x}=\bf{M}^T\bf{DM}+\bf{S}_{x0}$, where $\bf{D(\omega)}$ is a diagonal matrix with the spectral density for fluctuations in each of the six internal circuit parameters and $\bf{S_{x0}}$ is the white noise from the microwave readout amplifier.}

\redT{The remaining steps in the NEP calculation are: (i) compute the inverse of the cross-spectral density matrix \purple{$\bf{S_{x}^{-1}}$}, (ii) compute the vector signal expected from a \purple{unit impulsive} change in optical power \purple{$\bf{x}_{sig}(\omega)$}, (iii) compute the signal-to-noise ratio \purple{spectral }density \purple{$\sigma(\omega)=\bf{x}_{sig}^\dagger\bf{S_{x}^{-1}}\bf{x}_{sig}$}\grey{by taking the square of the signal vector relative to the inner product defined by the inverse cross-spectral density matrix}, and finally (iv) obtain \grey{the functional form of }the NEP from the \purple{square root of the }reciprocal of the signal-to-noise \purple{ratio }spectral density.} \grey{Note that NEP has units $\mathrm{W}/\sqrt(\mathrm{Hz})$ and the reciprocal of signal-to-noise has units of $1/\sqrt(\mathrm{Hz})$, but when NEP is normalized it has units that are comparable to the latter quantity.}

\begin{figure}
\includegraphics[width=0.45\textwidth]{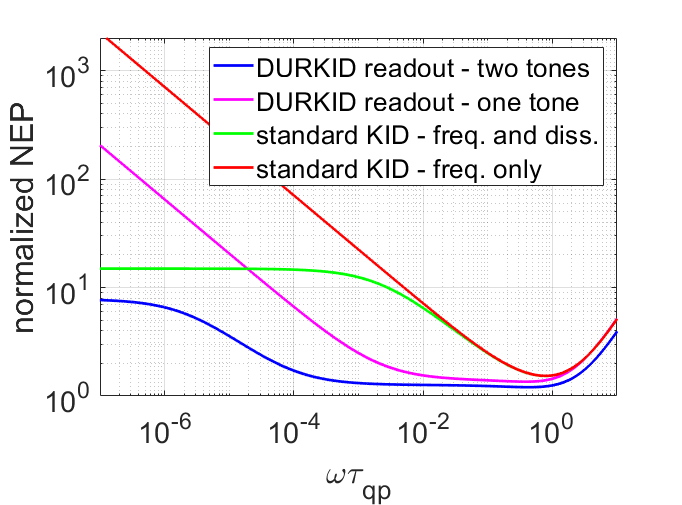}
\caption{\redT{Noise Equivalent Power (NEP) relative to the generation-recombination (G-R) noise level for 4 cases: a two-tone DuRKID readout (as intended), a single-tone DuRKID readout, and standard KID readout with the frequency quadrature readout, and a KID readout using both the frequency and dissipation quadrature (see main text for parameters). The horizontal axis shows signal frequency multiplied by the quasiparticle lifetime \purple{$\tau_{qp}$}. Note that the DuRKID read out with 2 tones (blue trace) performs substantially better in this case than the standard KID with 2 quadratures (green line), over a range of frequencies below the inverse quasiparticle lifetime. }
}
\label{NEP}
\end{figure}

\redT{Fig. \ref{NEP} shows the resulting NEP versus signal frequency for four example cases: (i) DuRKID readout using two-probe tones to monitor both modes of the dual-resonator, (ii) DuRKID readout using only a single probe tone, (iii) single-mode KID readout using both its frequency and dissipation response channels, and (iv) single-mode KID readout using only its frequency channel.  As discussed above, we chose to model the case of a dissipation-to-frequency responsivity ratio of 0.03.  Th\grey{e}\purple{is illustrative example} analysis uses a ratio of HEMT white noise power to generation-recombination (G-R) noise power at zero frequency of 0.2, a ratio of TLS noise power at $\omega_{QP}\purple{= 1/\tau_{qp}}$ to G-R noise power at zero frequency of 0.5, and a ratio of TLS fractional capacitance noise in $C_5$ at 2 times larger than that in the designed capacitors ($C_1 $--$ C_4,$). Additionally, the frequency dependence of the TLS noise is taken to be an ideal $1/f$ exponent. The NEP shown in the figure is normalized to (divided by) the G-R noise \purple{ at zero frequency}.}

\redT{The single-tone readout of the DuRKID shows the NEP degrading without limit as the signal frequency approaches zero (purple line). This limiting behavior occurs because, when only two components (i.e., the in-phase and quadrature (I-Q) response from the single probe tone) are measured, there exists a linear combination of TLS capacitance fluctuations in $C_5$, $C_2$ (or $C_4$), and $C_1$ (or $C_3$) that looks similar to a response from quasiparticle generation caused by weak optical illumination of the inductor (with fractional inductance change accompanied by the assumed $3\%$ smaller change in $1/Q$ due to quasiparticle dissipation). In contrast, when four components are measured (i.e., the I and Q components from $f_-$ and $f_\text{+}$), then the signal vector for a quasiparticle response to inductor illumination is not contained within the vector space of outputs in response to all possible changes in $C_5$, $C_2$ (or $C_4$), and $C_1$ (or $C_3$). In other words, if only a single-tone readout is used from the double resonator DuRKID circuit, then there is not enough information measured to distinguish an optical response from some possible combinations of TLS fluctuations in the various capacitors in the circuit. Also, we find that $C_5$ TLS fluctuations are not needed to spoil the single-tone readout of a DuRKID. Symmetry breaking in the circuit with a single-tone readout can give a nonzero response at $f_-$ to $C_1$ and $C_3$ fluctuations, and $f_\text{+}$ to $C_2$ and $C_4$ fluctuations, which can also spoil a DuRKID’s information and the related ability \purple{to }overcome TLS noise.}

\redT{However, when a two-tone readout is used in a DURKID (blue line), one can best distinguish an optical response from any possible TLS fluctuation. This contrasts \purple{with }a standard single-resonator KID where a single-tone I-Q readout does suffice to give an optical response signal vector that is outside the (1D) vector space of fluctuations in the readout. In a standard KID detector, the frequency-only quadrature measurement is generally worse (red line) than the one-tone DuRKID readout (purple line) due to less information about the signal, resulting in higher $1/f$ noise. In fact, the one-tone DURKID readout has an advantage in this comparison for frequencies $\omega\tau_{qp}<10^{-1}$ because the DuRKID distinguishes signal from $1/f$ noise due to new information of the signal relative to capacitance noise in this frequency regime. When the dissipation quadrature is added to the standard KID readout (green line), TLS noise is partially avoided and it performs within a factor of two of the intended DuRKID in NEP at the lowest frequencies. The most striking regime for improvement for the DuRKID is an intermediate frequency range from $10^{-5}<\omega\tau_{qp}<10^{-1}$. This behavior is related to the low influence from noise in $C_5$ from the DuRKID (blue line) relative to a standard KID, which can only mitigate $1/f$ noise above some noise level (green line). In an intermediate frequency regime, which of course depends on the model parameters, one can expect the DuRKID to perform as an improved alternative to the standard KID.}

\subsection{System Noise Discussion}

\redT{The NEP generally depends on the optical (absorber) power \cite{GEP} and readout power. Empirically, TLS-induced phase noise is proportional to the inverse root of measurement power \cite{NoisePropGao}. However, the generation-recombination noise will also increase with optical power. To allow a close comparison between a standard KID and a DuRKID, the DuRKID could be read out at approximately twice the total readout power relative to a standard KID such that the stored energy per inductor is equal in the KIDs. Furthermore, the reference KID should also use the same capacitor volume as the DuRKID. In this case, there will be a root-two decrease in RMS voltage in the DURKID relative to the KID. This will slightly increase the TLS $1/f$-noise for one measurement channel of the DuRKID relative to the KID. However, there is a potential order-of-magnitude decrease in NEP in the DuRKID. While the DuRKID performance can be substantially better than the standard KID, the TLS noise difference is not precisely known because there can be some decrease in noise from saturation of the TLSs by the adjacent readout tone. }

\red{Due to the power dependences of NEP, the DuRKID seems most likely to make an impact on low-illumination power IR KIDs. One such KID is a mid-IR KID designed for NASA's GEP mission \cite{GEP}. Another example is an IR KID detector that might be designed to reach the power sensitivity permitted by the Origins Space Telescope (OST), $10^{-20}$W$/\sqrt{Hz}$ \cite{OmidPhotonCounting}.}

\redT{The generation-recombination noise in the inductor of a KID relates to the breaking of Cooper pairs, which is a quasiparticle noise term \cite{FundamentalNoiseInKIDs}. Quasiparticle noise is quantitatively different in quantum information processing hardware than KID hardware for two reasons. Firstly, kinetic inductance is rarely used in a film-based inductor in quantum information science because there is a desire for low sensitivity to quasiparticles, which will induce loss in accord with the two-fluid model. Thus, a low-density quasiparticles is needed for low noise in the former, while high kinetic inductance in films \cite{GranAl} is especially needed in the latter for a substantial illumination response despite the finite background in quasiparticle noise. 
}

\gray{The most straightforward analysis of the system uses one feedline, e.g., ports 1 and 2. This corresponds to simulated data of $S_{21}$ shown in Fig. 5(c). The advantage of the method is sufficient for TLS noise distinction and in a future version of the DuRKID, one may choose to eliminate the second feedline. However, one could instead measure with 2 feedlines, where each feedline measures one resonator. In this case, when a signal perturbs the frequencies, each resonator signal could be set equal in magnitude as each mode takes on an emphasized character of one of the physical resonator modes. We also measured unintended leakage (crosstalk) between the two different feedlines at $|S_{14}|\sim 10^{-4}$ (this was measured with resonators detuned to see the stray coupling near either the A or B resonator mode). From this, we extracted the couplings of the resonators to the unintended feedlines as  $\kappa_{ax} = \kappa_{bx} \sim 20$ kHz  (see Appendix A).}  
 
\gray{We specified a TLS noise mitigation technique for ideal fabrication, where the hybridized mode noise is perfectly uncorrelated. However, noise sources can become correlated if the capacitor bridge is asymmetric, or if the inter-resonator coupling becomes too small to create a distinction between the modes. In practice, there will be some correlation between the components even in an optimal experimental circuit due to measurement noise and factors that are not addressed above. Thus in the future, one may want to construct an optimal linear filter based on the correlations and other noise mechanisms (white noise from amplifiers included, for example). In one approach for this, the optimal linear filter can be computed by measuring the 4-component noise and then computing the cross-spectral density matrix (see Eqs. 13 and 16 in Ref. \cite{Stevenson}).} 
\section{V. CONCLUSION}
\red{We have proposed a new KID detector design named DuRKID, which includes an electrical bridge and two resonance modes. The DuRKID was fabricated with two feedlines, where each is primarily coupled to one of the resonance modes when the modes are not hybridized. Measurements of the fabricated device revealed that the frequency tuning of one mode allowed the hybridization of the modes, as required for the intended operation. In the DuRKID, the TLS noise in the two modes is caused by different TLSs and capacitors such that the noise from the groups of TLSs is qualitatively different.}

\gray{In summary, we have proposed a new KID detector prototype consisting of two superconducting circuit resonators. We first described the fabricated device and schematic consisting of two resonators that both share an electrical capacitance bridge. We then showed the basic transmission and fittings of the device data along with frequency tuning of one mode which is necessary for device operation. $1/f$ noise will naturally arise from TLSs in the dielectric of the capacitors, as found in previous studies. In our study the capacitor configuration in the device allows the two resonators to share common TLSs generally, but this changes once the resonators are tuned to degeneracy, where a given TLS only perturbs one of the two hybridized resonator modes due to its residency in one of two TLS groups (n or m). This leads to uncorrelated noise in the two resonator modes since they experience separate noise sources. }

\red{From transmission measurements on the DuRKID, we found that the coupling between the modes is larger than the coupling of a TLS to a resonator mode. Furthermore, to achieve degeneracy we merely needed to tune a resonator with an applied dc-bias current, where the largest tuning was ~120 MHz. To show that a specific TLS is coupled to only one of the fully hybridized modes, we applied the input-output theory, a formalism from quantum information science, to analyze the DuRKID. This model involved 4 ports coupled to 2 resonators, with one representative TLS of each capacitor group in the bridge, and allowed a first model of TLS noise in the DuRKID. The model also allowed us to extract the capacitor loss tangents, inter-resonator coupling, and the intended and unintended couplings to transmission lines from data.}

\gray{We performed transmission measurements on the DuRKID and found experimentally that the coupling between the modes is larger than the coupling of a TLS to a resonator mode. Furthermore, to achieve degeneracy, we tuned the resonators by up to ~120 MHz with an applied current. 
To show that any given TLS will be coupled to only one of the two fully hybridized modes, we then analyzed the device schematic theoretically using a model with 4 ports coupled to 2 resonators, where a representative TLS of each TLS group resides in the bridge. The simulation shows that TLS noise is different than the intended KID signal. From the transmission data we obtain the TLS loss tangents, inter-resonator coupling, and the intended and unintended couplings to transmission lines.}

\red{From our first model of $1/f$ noise in the DURKID, we found that the TLS noise gives a lack of correlation in the two hybridized modes over time, as expected. This represents a potentially useful capability in a KID-type detector. However, we also modeled the system NEP of our DuRKID with a comparison to a standard KID using the expected noise sources and all of the possible readout channels. The DuRKID generally gives additional information, and thus lower NEP, relative to a standard KID. Moreover, the analysis shows \gray{not only that it lowered the noise, but} that there \red{can be a multi-decade wide} \gray{is a} frequency band that exhibits \red{over an order of magnitude lower noise}\gray{very low noise fluctuations} for a DuRKID relative to a standard KID. In the future, we plan to measure the NEP of a DuRKID as a function of frequency.}

\gray{In the future, we plan to demonstrate the uncorrelated noise in the device from TLSs. In principle, the measurement can be made by monitoring the lack of correlation over time while the resonator pair is near the fully hybridized state. We specify the TLS noise mitigation technique in terms of 2 quadratures of resonator transmission measurements of amplitude and phase at 2 frequencies (4 components) of the DuRKID. The noise from the hybridized state can be compared when the resonators are non-hybridized, where the noise from TLSs should appear as correlated noise on the resonators.}

\section{acknowledgments}
The authors thank F. C. Wellstood, C. Richardson, R. Ruskov and W. Wustmann for scientific discussions. This work was partially funded through a NASA Science Innovation Fund (SIF) award. B. Sarabi acknowledges funding through the Intelligence Community Postdoctoral Research Fellowship Program.

\appendix
\section*{Appendix A: Nonlinearity of the kinetic inductance and the fitting procedure}
\section{1. Nonlinearity of the kinetic inductance}
\begin{figure*}
\centering
\includegraphics[width=0.7\textwidth]{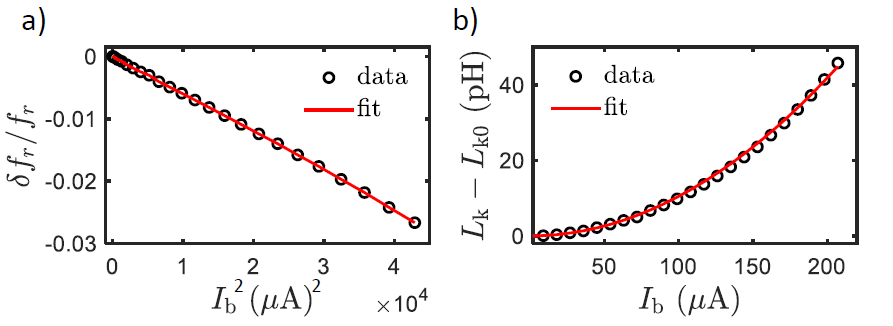}
\caption{a) Frequency and b) kinetic inductance dependence
on $I_{b}$.}   
 \label{fitinductance}
\end{figure*}

As described in the main text, the inductance of the LC resonators
is made tunable with DC bias. We measured the nonlinearity explicitly
in device 2 at milliKelvin temperature with a direct applied bias current (note that device 1 and 3 share a single dc-input current in parallel).
The relationship was given previously for inductance as a quadratic
function of bias current, but here we describe the nonlinear inductance.
The TiN inductor width is 5 mm wide.
Fig. \ref{fitinductance}(a) shows frequency of the resonator as a
function of the bias current squared. The fractional resonance frequency
shift can be written in terms of current or inductance as:
\textcolor{black}{
\begin{equation}
\frac{\delta f_{r}}{f_{r}}=-\frac{I_{b}^{2}}{2I'^{2}}=-\frac{\delta L_{r}}{2L_{r}}.
\end{equation}
}
We fit to a quadratic model as a function of bias current $I_{b}$
as shown in Fig. \ref{fitinductance}(a) and from this extracted $I'$=
789 $\mu$A. In Fig. \ref{fitinductance}(b) we show the same type of fit,
except using inductance as the independent variable rather than the
frequency.\\
\section{2. The fitting procedure}
\textcolor{black}{In order to fit the measured complex transmission
data, $S_{21}$, $S_{43}$ and~ $S_{31}$ we use Eqs. (\ref{eq:S21general}),
(\ref{eq:S43general}) and (\ref{eq:S41}). Although general formulas can
be made to directly fit for the many parameters simultaneously, such
multi-parameter fitting problems can be extremely sensitive to the
initial values due to the local optimization of the parameters, and
different amounts of noise in different data sets. Fitting a small
subset of the data leads to under-determined constraints and inaccuracies
due to covariances. To resolve this issue, we utilize an iterative
procedure through which our fitting problem, with 10 parameters, is
broken down into several independent fitting problems with carefully
chosen fixed parameters. Each fit contains only a few (two or three)
parameters at a time.}
First, we extract the cross-coupling $\varOmega_{AB}$ from the data of Fig. 3(c). We have previously performed resonator fittings from transmission data (e.g., $S_{21}$
) using the diameter correction method (DCM) (for a single resonator)
\cite{Khalil2012}. This technique uses a Least-Squares Monte Carlo
(LSM) method to find the minimum error, $\chi^{2}$.
To fit both $S_{21}$ and $S_{43}$ we developed
a fitting sequence using a sequence of different LSM fittings in multi-step
procedure. We start by using LSM fitting primarily with $S_{21}$
data, using the device model (equation (18)), except that an extra multiplicative
factor $e^{i\theta}$ is applied to the right-hand side of Eq. (18)
to account for the transmission line length. Initial guesses must
be placed for all variables, including $\omega_{A}$, $\omega_{B},$
$\kappa_{A}$, $\kappa_{B}$, $\gamma_{A}$, $\gamma_{B}$, $\kappa_{Ax}$,
$\kappa_{Bx}$ . Within the LSM fitting a Monte Carlo guess is made
in the form $x=x_{0}e^{\zeta\xi}$ for each fit parameter, where $x_{0}$
is the previous (initial) guess for the fit parameter $x$, $\zeta$
is a randomly generated number between -1 and 1, and $\xi$ is a parameter
smaller than one, which determines the MC guess domain.

However, to optimize $S_{21}$ in the first fit
we use two different $\xi$ for the different parameters. For parameters
associated with resonator B, we use fast iterations e.g. using $\xi=0.1$
for\textcolor{black}{{} $\omega_{A}$,} \textcolor{black}{$\kappa_{A}$,
$\gamma_{A}$ , $\kappa_{Ax}$, $\theta_{21}$ }to optimize it quickly.
Meanwhile, for parameters associated with resonator A, we use slow
iterations, e.g., using $\xi=10^{-5}$ for\textcolor{black}{{} $\omega_{B},$
$\kappa_{B}$, $\gamma_{B}$, $\kappa_{Bx}$}, \textcolor{black}{$\theta_{43}$
}, to leave the parameters with only slight changes. After this
step, the updated values are used to fit \textcolor{black}{$S_{43}$}
following the same procedure, with the difference that this time the
variation of \textcolor{black}{resonator A} parameters are quickly
optimized and resonator B parameters are mainly left unchanged.
Next the \textcolor{black}{$S_{43}$} fit parameters are determined,
such that \textcolor{black}{$S_{21}$} could be fit, and this completes
one cycle of our procedure. The previous 2 steps are repeated in
sequence: for example by using the fit results from $S_{21}$
($S_{43}$) we then fit $S_{43}$
($S_{21}$). The cycles of fitting are then repeated
iteratively for many times until convergence is obtained. The standard
resonator coupling to the intended transmission line is $Q_{c}^{A,B}=\frac{\omega_{A,B}}{2\kappa_{A,B}}$. The internal loss from one resonator is often desired, which is
$Q_{i}=\frac{\omega}{\gamma}$ in our devices, where the resonator
loss $1/Q_{i}$ is due to the sole dielectric in each resonator. Since
we have a dual transmission line setup, it is useful to define the apparent
internal quality factor for each resonator $Q_{i,app}^{A,B}=\frac{\omega_{A,B}}{\gamma_{A,B}+2\kappa_{Ax,Bx}}$,
which not only depends on the internal loss rate of the resonator
to its internal loss $\gamma_{a,b}$ , but also the coupling loss
to the unintended transmission line $2\kappa_{Ax,Bx}$, respectively.
\textcolor{black}{Figure 7} (a) and (b) show an example of such a fit to two resonance lineshapes for resonators A and B which are coupled to transmission lines 1-2 and 3-4 respectively for -140 dB power at device. The extracted fitting parameters from
$S_{21}$ and $S_{43}$ are used as an initial guess to fit the cross
transmission from port 1 to 4, $S_{41}$.
\begin{figure*}
\includegraphics[width=12cm]{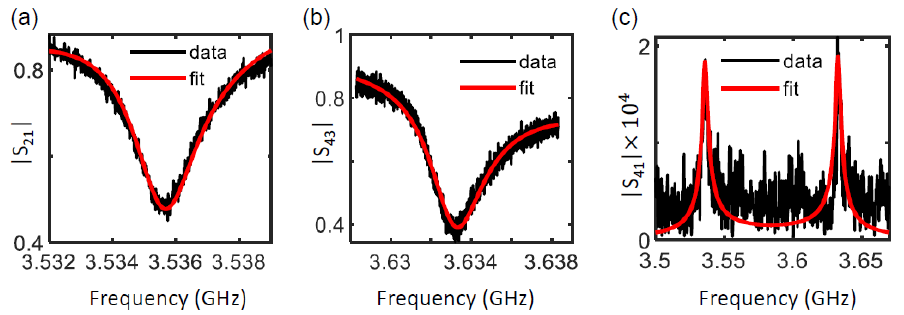}
\caption{\label{fig:Transmission-data-in}Transmission data in the absence
of bias for resonator A\textcolor{black}{{} (a) and resonator B (b).
The red curves show the fit of data to the model. The fit yielded $\omega_{A}=3.5357$
GHz, $\gamma_{A}=0.88$ MHz, $\kappa_{A}=0.9$ MHz, $\kappa_{Ax}=20$
kHz, and $\omega_{B}=3.6335$ GHz, $\gamma_{B}=0.77$ MHz, $\kappa_{B}=1.01$
MHz, $\kappa_{Bx}=19.7$ kHz, $Q_{i}^{A}=2741$, $Q_{i}^{B}=3026$
and $Q_{c}^{A}=1930$, $Q_{c}^{B}=1800$}. (c) Crosstalk coupling
(unintended coupling) between ports 1 and 4 of device 2 in the absence
of bias. Fitting data gives 20.3 kHz of cross-coupling between port
1 and 4 with\textcolor{black}{{} $\omega_{A}=3.534$ GHz, $\gamma_{A}=0.91$
MHz, $\kappa_{A}=0.97$ MHz, $\kappa_{Ax}=20.3$ kHz, and $\omega_{B}=3.6335$
GHz, $\gamma_{B}=0.77$ MHz, $\kappa_{B}=1.01$ MHz, $\kappa_{Bx}=19.5$
kHz.} Agreement between data and 4-port model confirms the appropriateness of the fit method and the many parameter models.}
\end{figure*}
\onecolumngrid
\section*{Appendix B: Derivation of the transmission for two resonances interacting with TLS}
\textbf{\textcolor{black}{1. LC resonator pair with an internal electrical bridge of capacitors}}\textbf{\large{} }{\large\par}
\medskip{}
The system we consider in Fig. 1(a) consists of two resonators and TLSs, described by the Hamiltonian $H_{sys}$ (Eq. 2). It is coupled to 2 transmission lines that are accessible at their ends via 4 ports. The system has resonator modes $a$ and $b$. According to the standard theory of input and output for quantum dissipative systems \cite{Gardiner}, the transmission lines can be modeled as a heat bath. Additionally, the couplings can be specified between transmission lines and the intended couplings, $\kappa_{A}$ and $\kappa_{B}$, as well as the unintended couplings, $\kappa_{Ax}$ and $\kappa_{Bx}$. The Heisenberg equation of motion for the bare resonator modes in terms of the system and input-output fields can be written as Eq. (2), (3),


\begin{eqnarray}
\frac{d}{dt}\left\langle a\right\rangle = &  & -i\omega\left\langle a\right\rangle =-\frac{i}{\hbar}\left\langle [a,H_{sys}]\right\rangle +\kappa_{A}\left\langle a\right\rangle -\sqrt{\kappa_{A}}\left\langle a_{1,out}\right\rangle -\sqrt{\kappa_{A}}\left\langle a_{2,out}\right\rangle \\
 &  & -\sqrt{\kappa_{Ax}}\left\langle b_{3,out}\right\rangle -\sqrt{\kappa_{Ax}}\left\langle b_{4,out}\right\rangle {\color{black}{\color{black}-\gamma_{A}\left\langle a\right\rangle +\kappa_{Ax}\left\langle a\right\rangle +\sqrt{\kappa_{A}\kappa_{Bx}}\left\langle b\right\rangle +\sqrt{\kappa_{Ax}\kappa_{B}}\left\langle b\right\rangle ,}}\nonumber 
\end{eqnarray}
and
\begin{eqnarray}
\frac{d}{dt}\left\langle b\right\rangle = &  & -i\omega\left\langle b\right\rangle =-\frac{i}{\hbar}\left\langle [b,H_{sys}]\right\rangle +\kappa_{B}\left\langle b\right\rangle -\sqrt{\kappa_{B}}\left\langle b_{3,out}\right\rangle -\sqrt{\kappa_{B}}\left\langle b_{4,out}\right\rangle \\
 &  & -\sqrt{\kappa_{Bx}}\left\langle a_{1,out}\right\rangle -\sqrt{\kappa_{Bx}}\left\langle a_{2,out}\right\rangle {\color{black}{\color{black}-\gamma_{B}\left\langle b\right\rangle +\kappa_{Bx}\left\langle b\right\rangle +\sqrt{\kappa_{B}\kappa_{Ax}}\left\langle a\right\rangle +\sqrt{\kappa_{Bx}\kappa_{A}}\left\langle a\right\rangle }}.\nonumber 
\end{eqnarray}
We write the equation of motion for the TLS operator $\sigma_{i}^{-}$ for TLS `i' using the Bloch equations approximation for relaxation and decoherence
\begin{eqnarray}
\frac{d}{dt}\left\langle \sigma_{i}^{-}\right\rangle = &  & -i\omega\left\langle \sigma_{i}^{-}\right\rangle =-i\omega_{TLS,i}\left\langle \sigma_{i}^{-}\right\rangle {\color{black}-\frac{\gamma_{TLS,i}}{2}\left\langle \sigma_{i}^{-}\right\rangle +2g\left\langle \sigma_{i}^{z}a\right\rangle +2g\left\langle \sigma_{i}^{z}b\right\rangle },
\end{eqnarray}
where the TLS decoherence rate is $\gamma_{TLS,i}=k_{1,i}+2k_{2,i}$, where typically $k_{1,i}=A_1$$\varDelta_{0,i}^{2}$coth ($\frac{\hbar\omega}{2k_{B}T}$)
and $k_{2,i}\approx A_2$T$^{2}$. The rates $k_{1,i}$ and $k_{2,i}$ describe TLS $i$ relaxation and phase decoherence rates associated with TLS-photon and TLS-TLS interactions, respectively and $A_1$ and $A_2$ are material-related constants \cite{Bahmanthesis}. In the low-temperature
and low-drive power limit ($k_{B}T\ll\hbar\omega$ and $n_{ph}\ll1$), the resonator will be asymptotically close to the ground state. This
suggests that we can replace the spin operator $\left\langle \sigma_{i}^{z}\right\rangle $
in Eq. (6) with the ground state value $-\frac{1}{2}$. With these
assumptions we obtain a closed system of linear equations that can
be solved for the TLS operators:
\begin{equation}
\left\langle \sigma_{i}^{-}\right\rangle =\frac{2g(\left\langle a\right\rangle +\left\langle b\right\rangle )\left\langle \sigma_{i}^{z}\right\rangle }{i(\omega_{TLS,i}-\omega)+\frac{\gamma_{TLS,i}}{2}}.\label{eq:sigmin}
\end{equation}
For thermally excited TLS, one can use a mean-field approach replacing
the operator $\left\langle \sigma^{z}_{i}\right\rangle $ in Eq. (6)
with its thermodynamic average value, i.e.. $\left\langle \sigma_{i}^{z}{}_{th}\right\rangle =-\frac{1}{2}\tanh(\frac{\hbar\omega}{2k_{B}T}).$
This approach is consistent with previous analysis of sound and microwave
absorption by TLSs \cite{Hunklinger1986, Classen1994}. The average
value of $\left\langle \sigma_{i}^{z}\right\rangle $ at a higher photon
regime is calculated in the next section.
\begin{figure}
\includegraphics[width=14cm]{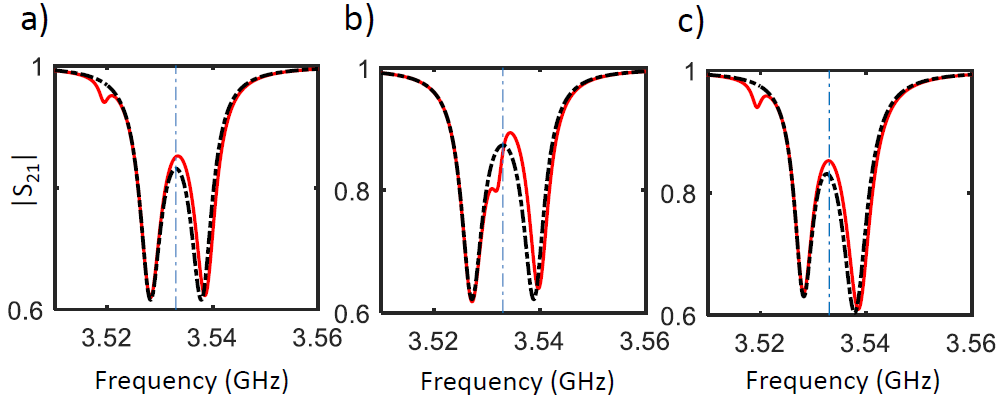}
\caption{\label{figcrosscoupling}a) Simulated transmission magnitude $|S_{21}|$ (with $Z_{B}\rightarrow0$ ) for a single TLS off
resonance coupled to a+b field in the bridge design. The TLS frequency
of $\omega_{tls}=3.52$ GHz is used here and the rest of the parameters
are the same as before. Note that dispersion is not caused at the
neighboring frequency mode, but at the mode that has the proper coupling
for the TLSs capacitor. b) The impact of a single TLS on resonance
coupled to a+b field on both hybridized modes. c) The impact of nonzero unintended coupling of each resonator to the other transmission
line, $\kappa_{Ax}=\kappa_{Bx}=20$kHz for data presented in (a).}
\end{figure}
To proceed, we now substitute the solution for $\left\langle \sigma_{i}^{-}\right\rangle $
(Eq. \ref{eq:sigmin}) into Eqs. (2-4) and obtain the solution for
resonator mode $a$ and $b$:
\begin{equation}
L_{A,+}\left\langle a\right\rangle +L_{AB,+}\left\langle b\right\rangle =\sqrt{\kappa_{A}}\left\langle a_{1,in}\right\rangle +\sqrt{\kappa_{A}}\left\langle a_{2,in}\right\rangle +\sqrt{\kappa_{Ax}}\left\langle b_{3,in}\right\rangle +\sqrt{\kappa_{Ax}}\left\langle b_{4,in}\right\rangle ,
\end{equation}
\begin{equation}
L_{A,-}\left\langle a\right\rangle +L_{AB,-}\left\langle b\right\rangle =-\sqrt{\kappa_{A}}\left\langle a_{1,out}\right\rangle -\sqrt{\kappa_{A}}\left\langle a_{2,out}\right\rangle -\sqrt{\kappa_{Ax}}\left\langle b_{3,out}\right\rangle -\sqrt{\kappa_{Ax}}\left\langle b_{4,out}\right\rangle ,
\end{equation}
\begin{equation}
L_{B,+}\left\langle b\right\rangle +L_{AB,+}\left\langle a\right\rangle =\sqrt{\kappa_{B}}\left\langle b_{3,in}\right\rangle +\sqrt{\kappa_{B}}\left\langle b_{4,in}\right\rangle +\sqrt{\kappa_{Bx}}\left\langle a_{1,in}\right\rangle +\sqrt{\kappa_{Bx}}\left\langle a_{2,in}\right\rangle ,
\end{equation}
\begin{equation}
L_{B,-}\left\langle b\right\rangle +L_{AB,-}\left\langle a\right\rangle =-\sqrt{\kappa_{B}}\left\langle b_{3,out}\right\rangle -\sqrt{\kappa_{B}}\left\langle b_{4,out}\right\rangle -\sqrt{\kappa_{Bx}}\left\langle a_{1,out}\right\rangle -\sqrt{\kappa_{Bx}}\left\langle a_{2,out}\right\rangle ,
\end{equation}
where,
\begin{equation}
L_{A,\pm}=-i\omega+i\omega_{A}+\gamma_{A}\pm\kappa_{A}\pm\kappa_{Ax}-\sum_{i=1}^{N}\frac{2sg_{i}^{2}}{i(\omega_{TLS,i}-\omega)+\frac{\gamma_{TLS,i}}{2}},
\end{equation}
\begin{equation}
L_{AB,\pm}=i\Omega_{AB}\pm\sqrt{\kappa_{A}\kappa_{Bx}}\pm\sqrt{\kappa_{Ax}\kappa_{B}}-\sum_{i=1}^{N}\frac{2sg_{i}^{2}}{i(\omega_{TLS,i}-\omega)+\frac{\gamma_{TLS,i}}{2}},
\end{equation}
\begin{equation}
L_{B,\pm}=-i\omega+i\omega_{B}+\gamma_{B}\pm\kappa_{B}\pm\kappa_{Bx}-\sum_{i=1}^{N}\frac{2sg_{i}^{2}}{i(\omega_{TLS,i}-\omega)+\frac{\gamma_{TLS,i}}{2}}.
\end{equation}
To study one coherent input, we then set $\left\langle a_{2,in}\right\rangle =\left\langle b_{4,in}\right\rangle =\left\langle b_{3,in}\right\rangle =0$
to calculate $S_{21}$ and $\left\langle a_{2,in}\right\rangle =\left\langle b_{4,in}\right\rangle =\left\langle a_{1,in}\right\rangle =0$
to calculate $S_{43}$. The solution for the resonator modes $a$
and $b$ takes the form
\begin{equation}
\left\langle a\right\rangle =\frac{\sqrt{\kappa_{A}}-\frac{L_{AB,+}\sqrt{\kappa_{Bx}}}{L_{B,+}}}{L_{A,+}-\frac{L_{AB,+}^{2}}{L_{B,+}}}\left\langle a_{1,in}\right\rangle ,
\end{equation}
and
\begin{equation}
\left\langle b\right\rangle =\frac{\sqrt{\kappa_{Bx}}}{L_{B,+}}\left\langle a_{1,in}\right\rangle -\frac{L_{AB,+}}{L_{B,+}}\frac{\sqrt{\kappa_{A}}-\frac{L_{AB,+}\sqrt{\kappa_{Bx}}}{L_{B,+}}}{L_{A,+}-\frac{L_{AB,+}^{2}}{L_{B,+}}}\left\langle a_{1,in}\right\rangle .
\end{equation}
Using the boundary conditions, $a_{1,(2),out}=\sqrt{\kappa_{A}}a-a_{2,(1),in}+\sqrt{\kappa_{Bx}}b$ and $b_{3,(4),out}=\sqrt{\kappa_{B}}b-b_{4,(3),in}+\sqrt{\kappa_{Ax}}a$, and Eq. (14) and
(15), we obtain \textcolor{black}{$S_{21}=\frac{\left\langle a_{2,out}\right\rangle }{\left\langle a_{1,in}\right\rangle }$,
$S_{43}=\frac{\left\langle b_{4,out}\right\rangle }{\left\langle b_{3,in}\right\rangle }$
and $S_{41}=\frac{\left\langle b_{4,out}\right\rangle }{\left\langle a_{1,in}\right\rangle }$ as}
\begin{align}
S_{21} & =1-\frac{\kappa_{A}}{-i(\omega-\omega_{A})+\kappa_{A}+\gamma_{A}-\sum_{i=1}^{N}\frac{2Sg_{i}^{2}}{-i(\omega-\omega_{TLS}^{i})+\frac{\gamma_{TLS}^{i}}{2}}-\frac{(i\Omega_{AB}-\sum_{i=1}^{N}(\frac{2g_{i}^{2}}{-i(\omega-\omega_{TLS}^{i})+\frac{\gamma_{TLS}^{i}}{2}}))^{2}}{-i(\omega-\omega_{B})+\kappa_{B}+\gamma_{B}-\sum_{i=1}^{N}\frac{2Sg_{i}^{2}}{-i(\omega-\omega_{TLS}^{i})+\frac{\gamma_{TLS}^{i}}{2}}}},\label{eq:S21general}
\end{align}
\begin{equation}
S_{43}=1-\frac{\kappa_{B}}{-i(\omega-\omega_{B})+\kappa_{B}+\gamma_{B}-\sum_{i=1}^{N}\frac{2Sg_{i}^{2}}{-i(\omega-\omega_{TLS}^{i})+\frac{\gamma_{TLS}^{i}}{2}}-\frac{(i\Omega_{AB}-\sum_{i=1}^{N}(\frac{2g_{i}^{2}}{-i(\omega-\omega_{TLS}^{i})+\frac{\gamma_{TLS}}{2}}))^{2}}{-i(\omega-\omega_{A})+\kappa_{A}+\gamma_{A}-\sum_{i=1}^{N}\frac{2Sg_{i}^{2}}{-i(\omega-\omega_{TLS}^{i})+\frac{\gamma_{TLS}}{2}}}},\label{eq:S43general}
\end{equation}
and
\begin{equation}
S_{41}=\frac{\sqrt{\kappa_{Bx}\kappa_{B}}}{L_{B,+}}-\frac{L_{AB,+}(L_{B,+}\sqrt{\kappa_{B}\kappa_{A}}-L_{AB,+}\sqrt{\kappa_{B}\kappa_{Bx})}}{L_{B,+}(L_{B,+}L_{A,+}-L_{AB,+}^{2})}+\frac{\sqrt{\kappa_{Ax}\kappa_{A}}-\frac{\sqrt{\kappa_{Ax}}L_{AB,+}\sqrt{\kappa_{Ax}\kappa_{Bx}}}{L_{B,+}}}{L_{A,+}-\frac{L_{AB,+}^{2}}{L_{B,+}}}.\label{eq:S41}
\end{equation}
Reducing the calculation to equal coupling and photon decay rates
for each resonator: $\kappa_{A}=\kappa_{B}=\kappa_{r}$, $\gamma_{A}=\gamma_{B}=\gamma_{r}$,
and assuming no (cross) coupling to the unintended transmission line,
$\kappa_{Ax}=\kappa_{Bx}=0$, the transmission properties of this
four-port system coupling to a single TLS through the $a+b$ field
is given by
\begin{equation}
S_{43}=S_{21}=1-\kappa_{A}\frac{-i(\omega-\omega'_{r})+\kappa_{r}+\gamma_{r}}{-(\omega-\omega'_{r})^{2}+(\kappa_{r}+\gamma_{r})^{2}-(i\Omega_{AB}-\frac{\tanh(\frac{\hbar\omega}{2k_{B}T})g^{2}}{-i(\omega-\omega_{TLS})+\frac{\gamma_{TLS}}{2}})^{2}-2i(\kappa_{r}+\gamma_{r})(\omega-\omega'_{r})},
\end{equation}
where $\omega'_{r}=\omega_{r}-\frac{g^{2}}{\omega-\omega_{TLS}+i\frac{\gamma_{TLS}}{2}}.$
Similarly, coupling to a single TLS through the $a-b$ field can be obtained.
%
\\

\textbf{\textcolor{black}{2. Standard-coupled two-resonator}}\textbf{\large{} }{\large\par}
\medskip{}
\textcolor{black}{Considering the standard-coupled two-resonator}
with Hamiltonian of 
\begin{align}
H_{sys} & =\hbar\omega_{A}a^{\dagger}a+\hbar\omega_{B}b^{\dagger}b+\hbar\varOmega_{AB}(b^{\dagger}a+a^{\dagger}b)+\varepsilon_{TLS}\sigma^{z}\label{eq:standardH}\\
 & -i\hbar g_{A}(\sigma_{A}^{+}a+\sigma_{A}^{-}a^{\dagger})-i\hbar g_{B}(\sigma_{B}^{+}b+\sigma_{B}^{-}b^{\dagger}),
\end{align}
and assuming $g_{A}=g_{B}=g,$ we obtain 
\begin{equation}
S_{43}=S_{21}=1-\kappa_{A}\frac{-i(\omega-\omega'_{r})+\kappa_{r}+\gamma_{r}}{-(\omega-\omega'_{r})^{2}+(\kappa_{r}+\gamma_{r})^{2}+\Omega_{AB}^{2}-2i(\kappa_{r}+\gamma_{r})(\omega-\omega'_{r})}.
\end{equation}
\textbf{3. Coherent drive in the Hamiltonian (Treatment of the many-photon
case)}
\medskip{}

We consider the case of incoming coherent radiation from the left-hand
side of the transmission line, port 1 or 3. The presence of such coherent
drive is accounted for by an effective Hamiltonian
\begin{equation}
H=H_{sys}+H_{d},
\end{equation}
where the drive appears as a term $H_{d}=i\hbar Ja^{\dagger}+H.c.$, and $J=\sqrt{\Gamma_{ext}^{a}}A(t),$ and $\Gamma_{ext}^{a}=\kappa_{A}.$
This form is derived by assuming $\left\langle a_{1,in}(t)\right\rangle =A(t)$
and $\left\langle a_{2,in}(t)\right\rangle =\left\langle b_{3,in}(t)\right\rangle =\left\langle b_{4,in}(t)\right\rangle =0$.
The interaction of a single TLS with the resonators can be described
by the Jaynes-Cummings Hamiltonian in a frame rotating at the pump
frequency $\omega_{p}$ :
\begin{align}
H & =\hbar(\omega_{p}-\omega_{A})a^{\dagger}a+\hbar(\omega_{p}-\omega_{B})b^{\dagger}b+\frac{\hbar(\omega_{p}-\omega_{TLS})}{2}\sigma_{z}\nonumber \\
 & +\hbar\varOmega_{AB}(b^{\dagger}a+a^{\dagger}b)-i\hbar g(\sigma^{-}(a^{\dagger}+b^{\dagger})-\sigma^{+}(a+b))+i\hbar J(a^{\dagger}-a)\nonumber \\
\end{align}
The corresponding dissipation can be described by the Lindblad master equation \cite{Capelle2020}:
\begin{align}
\frac{d\rho}{dt} & =-i\hbar[H_{sys},\rho]+\Gamma_{\downarrow\uparrow}(n_{th}+1)\mathcal{D}_{\sigma}(\rho)+\frac{\Gamma_{\phi}}{2}\mathcal{D}_{\sigma z}(\rho)\nonumber \\
 & +\Gamma_{\downarrow\uparrow}n_{th}\mathcal{D}_{\sigma^{\dagger}}(\rho)+\Gamma_{ext}^{a}\mathcal{D}_{a}(\rho)+\Gamma_{ext}^{b}\mathcal{D}_{b}(\rho),
\end{align}
where the occupation number of TLS is $n_{th}=1/(e^{\hbar\omega/kT}-1)$, the damping of resonator A and B in the absence of TLS is $\Gamma_{ext}^{a}$=$\kappa_{A}$
and $\Gamma_{ext}^{b}=\kappa_{B}$, the TLS dephasing rate is $\Gamma_{\phi}$, its rate at zero temperature $\Gamma_{\downarrow\uparrow}$ , and \textcolor{black}{$\mathcal{D}_{A}(\rho)=A\rho A^{\dagger}-\frac{1}{2}(A^{\dagger}A\rho+\rho A^{\dagger}A).$}
Using $\left\langle A\right\rangle =Tr\left\langle A\rho\right\rangle $
and $(d/dt)\left\langle A\right\rangle =Tr\left\langle A(d/dt)\rho\right\rangle $,
we can compute the Maxwell-Bloch equations
\begin{equation}
\frac{d\left\langle a\right\rangle }{dt}=(-i(\omega_{p}-\omega_{a})-\frac{\Gamma_{ext}^{a}}{2})\left\langle a\right\rangle +g_{a}\left\langle \sigma\right\rangle +J
\end{equation}
\begin{equation}
\frac{d\left\langle b\right\rangle }{dt}=(-i(\omega_{p}-\omega_{b})-\frac{\Gamma_{ext}^{b}}{2})\left\langle b\right\rangle +g_{b}\left\langle \sigma\right\rangle 
\end{equation}
\begin{equation}
\frac{d\left\langle \sigma\right\rangle }{dt}=(-i(\omega_{p}-\omega_{TLS})-\Gamma_{2})\left\langle \sigma\right\rangle +g_{a}\left\langle a\sigma_{z}\right\rangle +g_{b}\left\langle b\sigma_{z}\right\rangle 
\end{equation}
\begin{equation}
\frac{d\left\langle \sigma_{z}\right\rangle }{dt}=-2g_{a}(\left\langle a^{\dagger}\sigma\right\rangle +\left\langle a\sigma^{\dagger}\right\rangle )-2g_{b}(\left\langle b^{\dagger}\sigma^{-}\right\rangle +\left\langle b\sigma^{+}\right\rangle )-\Gamma_{1}(\left\langle \sigma_{z}\right\rangle -\left\langle \sigma_{z}\right\rangle _{th}),
\end{equation}
where we define $\Gamma_{2}=(\Gamma_{\downarrow\uparrow}/2)(1+2n_{th})$+$\Gamma_{\phi},$
$\Gamma_{1}=\Gamma_{\downarrow\uparrow}(1+2n_{th})$, $\left\langle \sigma_{z}\right\rangle _{th}=-1/(1+2n_{th})=-$tanh
($\hbar\omega/2k_{B}T$). 
To transform this system Eqs. (28)-(31) into a closed set of equations, we
neglect the correlations and factorize the products $\left\langle a\sigma_{z}\right\rangle =$$\left\langle a\right\rangle \left\langle \sigma_{z}\right\rangle $,
$\left\langle a^{\dagger}\sigma_{z}\right\rangle =$$\left\langle a^{\dagger}\right\rangle \left\langle \sigma_{z}\right\rangle $,
$\left\langle b\sigma_{z}\right\rangle =$$\left\langle b\right\rangle \left\langle \sigma_{z}\right\rangle $,
$\left\langle b^{\dagger}\sigma_{z}\right\rangle =$$\left\langle b^{\dagger}\right\rangle \left\langle \sigma_{z}\right\rangle $.
We first determine the solution for the cavity field, $\left\langle a\right\rangle $
and $\left\langle b\right\rangle $, using the approximation $\left\langle a\right\rangle $=$\alpha+\delta\alpha(t)e^{-i(\omega_{p}-\omega_{a})t}$, $\left\langle b\right\rangle $=$\beta+\delta\beta(t)e^{-i(\omega_{p}-\omega_{b})t}, $$\left\langle \sigma\right\rangle $=$\sigma_{0}+\delta\sigma(t)e^{-i(\omega_{p}-\omega_{TLS})t}$
and $\left\langle \sigma_{z}\right\rangle =\sigma_{z0},$ where $\delta\alpha(t)$,
$\delta\beta(t)$ and $\delta\sigma(t)$ are slowly varying complex
functions \cite{Capelle2020}. The equation for the stationary components
are:
\begin{equation}
0=(-i(\omega_{p}-\omega_{a})-\frac{\Gamma_{ext}^{a}}{2})\alpha+g_{a}\sigma_{0}-i\Omega_{AB}\beta+J
\end{equation}
\begin{equation}
0=(-i(\omega_{p}-\omega_{b})-\frac{\Gamma_{ext}^{b}}{2})\beta+g_{b}\sigma_{0}-i\Omega_{AB}\alpha
\end{equation}
\begin{equation}
0=(-i(\omega_{p}-\omega_{TLS})-\Gamma_{2})\sigma_{0}+g_{a}\alpha\sigma_{z0}+g_{b}\beta\sigma_{z0}\label{eq:sigma0}
\end{equation}
\begin{equation}
0=-2g_{a}(\alpha^{*}\sigma_{0}+\alpha\sigma_{0}^{*})-2g_{b}(\beta^{*}\sigma_{0}+\beta\sigma_{0}^{*})-\Gamma_{1}(\sigma_{z0}-\left\langle \sigma_{z}\right\rangle _{th}).
\end{equation}
Grouping terms by terms related to $\alpha$ and $\beta$, one finds
\begin{equation}
\sigma_{0}=\frac{g_{a}\alpha\sigma_{z0}+g_{b}\beta\sigma_{z0}}{i(\omega_{p}-\omega_{TLS})+\Gamma_{2}}.
\end{equation}
The solution for the outfield is obtained from the boundary conditions
$a_{1,(2),out}=\sqrt{\Gamma_{ext}^{a}}a-a_{2,(1),in}$ with inserting
the assumed form of the input field $\left\langle a_{1,in}(t)\right\rangle =A(t)$.
From Eq. (33) and replacing $\sigma_{0}$ from (36) we have
\begin{equation}
0=(-i(\omega_{p}-\omega_{b})-\frac{\Gamma_{ext}^{b}}{2})\beta+\frac{g_{b}g_{a}\alpha\sigma_{z0}+g_{b}g_{b}\beta\sigma_{z0}}{i(\omega_{p}-\omega_{TLS})+\Gamma_{2}}-i\Omega_{AB}\alpha
\end{equation}
\begin{equation}
0=(-i(\omega_{p}-\omega_{b})-\frac{\Gamma_{ext}^{b}}{2}+\frac{g_{b}g_{b}\sigma_{z0}}{i(\omega_{p}-\omega_{TLS})+\Gamma_{2}})\beta+\frac{g_{b}g_{a}\alpha\sigma_{z0}}{i(\omega_{p}-\omega_{TLS})+\Gamma_{2}}-i\Omega_{AB}\alpha
\end{equation}
\begin{equation}
-(-i(\omega_{p}-\omega_{b})-\frac{\Gamma_{ext}^{b}}{2}+\frac{g_{b}g_{b}\sigma_{z0}}{i(\omega_{p}-\omega_{TLS})+\Gamma_{2}})\beta=\frac{g_{b}g_{a}\alpha\sigma_{z0}}{i(\omega_{p}-\omega_{TLS})+\Gamma_{2}}-i\Omega_{AB}\alpha
\end{equation}
\begin{equation}
-(-i(\omega_{p}-\omega_{b})-\frac{\Gamma_{ext}^{b}}{2}+\frac{g_{b}g_{b}\sigma_{z0}}{i(\omega_{p}-\omega_{TLS})+\Gamma_{2}})\beta=(\frac{g_{b}g_{a}\sigma_{z0}}{i(\omega_{p}-\omega_{TLS})+\Gamma_{2}}-i\Omega_{AB})\alpha.
\end{equation}
Finally we obtain
\begin{equation}
\beta=\frac{-\frac{g_{b}g_{a}\sigma_{z0}}{i(\omega_{p}-\omega_{TLS})+\Gamma_{2}}+i\Omega_{AB}}{(-i(\omega_{p}-\omega_{b})-\frac{\Gamma_{ext}^{b}}{2}+\frac{g_{b}g_{b}\sigma_{z0}}{i(\omega_{p}-\omega_{TLS})+\Gamma_{2}})}\alpha
\end{equation}
To proceed to the output fields, we next substitute the solution for
$\beta$ from Eq. (41) into Eq. (32)
\begin{equation}
0=(-i(\omega_{p}-\omega_{a})-\frac{\Gamma_{ext}^{a}}{2}+\frac{g_{a}g_{a}\sigma_{z0}}{i(\omega_{p}-\omega_{TLS})+\Gamma_{2}})\alpha+(-i\Omega_{AB}+\frac{g_{a}g_{b}\sigma_{z0}}{i(\omega_{p}-\omega_{TLS})+\Gamma_{2}})\beta+J
\end{equation}
\begin{equation}
(i(\omega_{p}-\omega_{a})+\frac{\Gamma_{ext}^{a}}{2}-\frac{g_{a}g_{a}\sigma_{z0}}{i(\omega_{p}-\omega_{TLS})+\Gamma_{2}})\alpha+(i\Omega_{AB}-\frac{g_{a}g_{b}\sigma_{z0}}{i(\omega_{p}-\omega_{TLS})+\Gamma_{2}})\beta=J
\end{equation}
\begin{equation}
(i(\omega_{p}-\omega_{a})+\frac{\Gamma_{ext}^{a}}{2}-\frac{g_{a}g_{a}\sigma_{z0}}{i(\omega_{p}-\omega_{TLS})+\Gamma_{2}})\alpha+(i\Omega_{AB}-\frac{g_{a}g_{b}\sigma_{z0}}{i(\omega_{p}-\omega_{TLS})+\Gamma_{2}})\frac{-\frac{g_{b}g_{a}\sigma_{z0}}{i(\omega_{p}-\omega_{TLS})+\Gamma_{2}}+i\Omega_{AB}}{(-i(\omega_{p}-\omega_{b})-\frac{\Gamma_{ext}^{b}}{2}+\frac{g_{b}g_{b}\sigma_{z0}}{i(\omega_{p}-\omega_{TLS})+\Gamma_{2}})}\alpha = J
\end{equation}
\begin{equation}
[(i(\omega_{p}-\omega_{a})+\frac{\Gamma_{ext}^{a}}{2}-\frac{g_{a}g_{a}\sigma_{z0}}{i(\omega_{p}-\omega_{TLS})+\Gamma_{2}})+\frac{(i\Omega_{AB}-\frac{g_{a}g_{b}\sigma_{z0}}{i(\omega_{p}-\omega_{TLS})+\Gamma_{2}})(i\Omega_{AB}-\frac{g_{a}g_{b}\sigma_{z0}}{i(\omega_{p}-\omega_{TLS})+\Gamma_{2}})}{(-i(\omega_{p}-\omega_{b})-\frac{\Gamma_{ext}^{b}}{2}+\frac{g_{b}g_{b}\sigma_{z0}}{i(\omega_{p}-\omega_{TLS})+\Gamma_{2}})}]\alpha=\sqrt{\Gamma_{ext}^{a}}a_{1,in}.
\end{equation}
The solution for the resonator $\alpha$ field is 
\begin{equation}
\alpha=\frac{\sqrt{\Gamma_{ext}^{a}}}{[(i(\omega_{p}-\omega_{a})+\frac{\Gamma_{ext}^{a}}{2}-\frac{g_{a}g_{a}\sigma_{z0}}{i(\omega_{p}-\omega_{TLS})+\Gamma_{2}})+\frac{(i\Omega_{AB}-\frac{g_{a}g_{b}\sigma_{z0}}{i(\omega_{p}-\omega_{TLS})+\Gamma_{2}})^{2}}{(-i(\omega_{p}-\omega_{b})-\frac{\Gamma_{ext}^{b}}{2}+\frac{g_{b}g_{b}\sigma_{z0}}{i(\omega_{p}-\omega_{TLS})+\Gamma_{2}})}]}a_{1,in}.
\end{equation}
Using this result one can find transmission and reflection coefficient using the boundary condition $a_{2,out}=\sqrt{\Gamma_{ext}^{a}}a-a_{1,in}$.
The transmission, $S_{21}=\frac{a_{2,out}}{a_{1,in}}$ can be written
as
\begin{equation}
S_{21}=1-\frac{\Gamma_{ext}^{a}}{[(i(\omega_{p}-\omega_{a})+\frac{\Gamma_{ext}^{a}}{2}-\frac{g_{a}g_{a}\sigma_{z0}}{i(\omega_{p}-\omega_{TLS})+\Gamma_{2}})+\frac{(i\Omega_{AB}-\frac{g_{a}g_{b}\sigma_{z0}}{i(\omega_{p}-\omega_{TLS})+\Gamma_{2}})^{2}}{(-i(\omega_{p}-\omega_{b})-\frac{\Gamma_{ext}^{b}}{2}+\frac{g_{b}g_{b}\sigma_{z0}}{i(\omega_{p}-\omega_{TLS})+\Gamma_{2}})}]},
\end{equation}
where 
\begin{equation}
\sigma_{z0}=\left\langle \sigma_{z}\right\rangle _{th}[1-\frac{\Gamma_{2}^{2}\overline{n}/n_{s}}{(\omega_{TLS}-\omega_{p})+\Gamma_{2}^{2}(1+\overline{n}/n_{s})}],
\end{equation}
and $\overline{n}=$$|A(t)|^{2}$ is the mean photon number in the
cavity and $n_{s}^{-1}=4g^{2}/\Gamma_{1}\Gamma_{2}$ the number of
photons required to saturate the TLS transition.
\\

\textbf{\textcolor{black}{3. Correlation amplitude:}}
\begin{figure*}
\includegraphics[width=10cm]{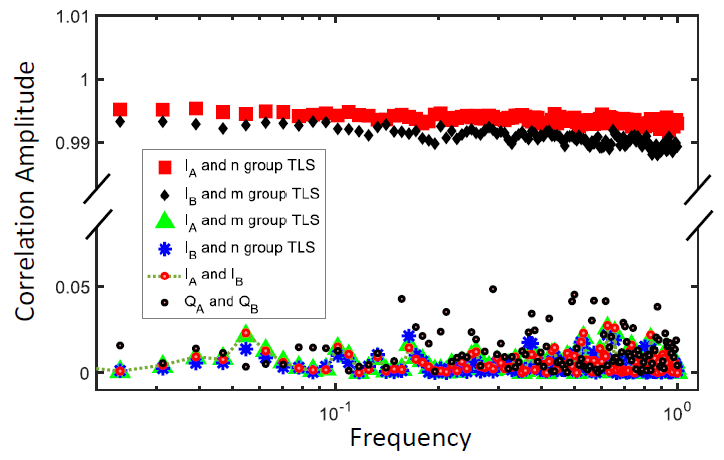}
\caption{Correlation amplitude $C_{xy}$ between noise quantities: $1/f$-noise TLS frequencies and resonator transmission quadratures ($I_A$, $Q_A$, $I_B$ and $Q_B$). $1/f$ noise sources from m-group TLS and n-group TLS are not correlated as there is no significant TLS-TLS coupling mechanism. There is near perfect noise correlation between $I_A$ and n-group TLS, and $I_B$ and m-group TLS. There is no significant normalized correlation between resonator quadratures $I_A$ and $I_B$ because the noise fluctuations occur as an m- or n- n-group TLS source on a corresponding resonator quadrature $I_A$ or $I_B$. This implies that noise is not significantly correlated through the resonator-resonator coupling. $Q_A$ and $Q_B$ have slightly larger normalized correlation amplitude, but this is of no consequence because these quadratures have small noise powers.}
\end{figure*}
\\
TLS ($1/f$) frequency noise causes noise in resonator transmission (e.g.
$S_{21})$, where the transmission changes in phase and amplitude. The noise is mainly seen in a particular
direction in IQ space as fluctuations in [Re $(S_{21} (f_{0A}))$, Im $(S_{21}(f_{0A}))]$ and [Re $(S_{21} (f_{0B}))$, Im $(S_{21}(f_{0B}))]$ which are tangent to the complex transmission plot, which we use (Eq. (18)) to define the quadratures $[Q'_{A}, I'_{A}]$ and $[Q'_{B}, I'_{B}]$ shown in Fig. 5(c). As in a standard KID, the fluctuation quadrature is thus a frequency noise. However, now we have two input frequencies and two TLS noise sources. The TLSs induce noise in the resonators, where $I'_{A}$ and $I'_{B}$ in Fig. 5(c) are defined to be tangent to the transmission versus frequency plot near the probe frequencies [$f_A$, $f_B$]. In Fig. 9 we show the correlation amplitude (the normalized cross-spectral density) 
\begin{equation}
C_{xy}=\frac{S_{\nu}^{xy}}{(S_{\nu}^{xx}S_{\nu}^{yy})^{1/2}},
\end{equation}
between different quadratures and TLS groups. Here, x and y are the different possible aforementioned TLS and quadrature noise types.  $S_{\nu}^{xy}$ is the cross-spectral density between x and y, and $S_{\nu}^{xx}$ and $S_{\nu}^{yy}$ are the autospectral density of x and y respectively.
The result plotted in Fig. 9 shows that the $1/f$ noise on each mode
and its own TLS fluctuations are highly correlated, but the noise induced on the measurement quadratures of the two resonator modes is uncorrelated.


\end{document}